\documentclass[12pt]{article}%
\usepackage[nosort]{cite}
\usepackage{graphicx}
\usepackage{multicol}
\usepackage{amsfonts}
\usepackage{amssymb}
\usepackage{amsmath}
\usepackage{heck}
\usepackage{afterpage}
\usepackage{setspace}
\usepackage{verbatim}
\usepackage{color}
\usepackage{longtable}
\usepackage{float}
\usepackage{subcaption}
\usepackage{epsfig}
\usepackage{enumerate}
\usepackage{epstopdf}
\usepackage[enableskew, vcentermath]{youngtab}
\usepackage{adjustbox}
\usepackage{multirow}
\usepackage{tikz}
\usepackage[margin=1in]{geometry}
\usepackage{titletoc}
\usepackage{rotating}
\usepackage{hyperref}
\usepackage{mathtools}%
\providecommand{\U}[1]{\protect\rule{.1in}{.1in}}
\pdfoutput=1
\newsavebox{\mysavebox}

\hypersetup{colorlinks,citecolor=black,filecolor=black,linkcolor=black,urlcolor=black}
\usetikzlibrary{decorations.markings}

\numberwithin{equation}{section}
\def\bZ{\mathbb{Z}}

\usetikzlibrary{chains}
\allowdisplaybreaks
\tikzset{node distance=2em, ch/.style={circle,draw,on chain,inner sep=2pt},chj/.style={ch,join},every path/.style={shorten >=4pt,shorten <=4pt},line width=1pt,baseline=-1ex}

\newcommand{\ba}{\begin{eqnarray}}
\newcommand{\ea}{\end{eqnarray}}

\newcommand{\cC}{\mathcal{C}}

\newcommand{\cK}{\mathcal{K}}

\newcommand{\cS}{\mathcal{S}}

\newcommand{\cW}{\mathcal{W}}
\newcommand{\centK}{C_\cC(\cK)}
\newcommand{\rhoend}{\rho_{\text{end}}}
\newcommand{\Kend}{\cK_{\text{end}}}

\newcommand{\centKend}{C_\cC(\cK_{\text{end}})}

\newcommand{\be}{\begin{equation}}
\newcommand{\ee}{\end{equation}}
\tikzstyle{startstop} = [rectangle, rounded corners, minimum width=3cm, minimum height=1cm,text centered, draw=black, fill=blue!10]
\tikzstyle{startstop} = [rectangle, rounded corners, minimum width=3cm, minimum height=1cm,text centered, draw=black, fill=blue!10]
\tikzstyle{io} = [trapezium, trapezium left angle=70, trapezium right angle=110, minimum width=3cm, minimum height=1cm, text centered, draw=black, fill=blue!30]
\tikzstyle{process} = [rectangle, minimum width=3cm, minimum height=1cm, text centered, draw=black, fill=orange!30]
\tikzstyle{decision} = [diamond, minimum width=3cm, minimum height=1cm, text centered, draw=black, fill=green!30]
\tikzstyle{arrow} = [thick,->,>=stealth]
\tikzset{->-/.style={decoration={
  markings,
  mark=at position #1 with {\arrow[scale=2.4]{>}}},postaction={decorate}}}
\makeatletter \@addtoreset{equation}{section} \makeatother

\begin{document}

\date{\today}

\title{Topological Operators and Completeness of Spectrum in Discrete Gauge Theories}

\institution{IAS}{\centerline{School of Natural Sciences, Institute for Advanced Study, Princeton, NJ 08540, USA}}

\authors{Tom Rudelius\footnote{e-mail: {\tt rudelius@ias.edu}}
and Shu-Heng Shao\footnote{e-mail: {\tt shao@ias.edu}}}

\abstract{In many gauge theories, the existence of particles in every representation of the gauge group (also known as completeness of the spectrum) is equivalent to the absence of one-form global symmetries.  
However, this relation does not hold, for example, in the gauge theory of non-abelian finite groups.   
We refine this statement by considering topological operators that are not necessarily associated with any global symmetry.  
For discrete gauge theory in three spacetime dimensions, we show that completeness of the spectrum is equivalent to the absence of certain Gukov-Witten topological operators.  We further extend our analysis to four and higher spacetime dimensions.  
Since topological operators are natural generalizations of global symmetries, we discuss evidence for their absence in a consistent theory of quantum gravity. 
}

\maketitle

\setcounter{tocdepth}{3}

\tableofcontents


\newpage

\section{Introduction \label{sec:INTRO}}

The last few years have seen great advances in the understanding of/appreciation for symmetries in quantum field theory and quantum gravity. In quantum field theory in $d$ spacetime dimensions, symmetries come in two varieties: gauge and global.  
 They are distinguished by an integer $0 \leq q \leq d-1$, which indicates the dimension of the manifolds on which operators charged under the symmetry are supported. Such a symmetry is referred to as a $q$-form symmetry, with $q=0$ an ordinary symmetry and $q > 0$ referred to as a \textit{higher-form symmetry} \cite{Gaiotto:2014kfa}.\footnote{See also \cite{Pantev:2005zs,Pantev:2005rh,Hellerman:2006zs,Caldararu:2007tc} for earlier discussions on higher-form symmetries in two spacetime dimensions.} They are associated with a group $G$, which may be either discrete or continuous. For an ordinary symmetry, $G$ may be either abelian or non-abelian, but for a higher-form symmetry $G$ is necessarily abelian.

A $q$-form global symmetry is generated by its \emph{symmetry operators}, which are topological operators $U_g(M^{(d-q-1)})$ that have support on manifolds $M^{(d-q-1)}$ of dimension $d-q-1$ and are labeled by elements $g$ of the group $G$. Here, \emph{topological} means that correlation functions involving the operator $U_g(M^{(d-q-1)})$ are invariant under continuous deformations of $M^{(d-q-1)}$ provided that $M^{(d-q-1)}$ does not intersect any other operators in the deformation process. Two such operators supported on the same manifold $M^{(d-q-1)}$ may be fused according to the group multiplication law to produce a third operator:
\begin{equation}\label{introgroup}
U_g(M^{(d-q-1)}) \times U_{g'}(M^{(d-q-1)}) = U_{g''}(M^{(d-q-1)})\,,~~~g g' = g''.
\end{equation}

Standard lore holds that global symmetries are not allowed in quantum gravity. Continuous zero-form global symmetries lead to stable black hole remnants \cite{Misner, Banks:2010zn}, which have been argued to cause problems for quantum gravity \cite{Susskind:1995da}. In string theory, any continuous zero-form global symmetry of the string worldsheet is gauged in the bulk spacetime \cite{Banks:1988yz}. Likewise in AdS$_d$/CFT$_{d-1}$, under certain assumptions, it has been argued that $q$-form global symmetries with $q \leq d-3$ (both continuous and discrete) of the boundary CFT are necessarily gauged in the bulk \cite{Harlow:2018tng}.

A second bit of quantum gravity lore is the \emph{completeness hypothesis}, which holds that any gauge theory coupled to quantum gravity must have objects charged under every finite-dimensional irreducible representation of the gauge group \cite{Polchinski:2003bq}. Such objects may correspond to individual particles or collections of multiple particles, but in this work we will abuse terminology slightly and refer to all such objects simply as ``particles.'' 
A gauge theory that satisfies this criterion is said to be \emph{complete}, whereas a gauge theory that violates the criterion is \emph{incomplete}. Arguments for the completeness of zero-form gauge theories have been given on the basis of black hole arguments \cite{Banks:2010zn} and in the context of of AdS/CFT \cite{Harlow:2018tng}.  See also \cite{Craig:2018yvw} for discussions in the case of discrete gauge theory. 

Finally, a third bit of lore holds that the first two are related to one another: a quantum field theory with gauge group $G$ will have an electric one-form global symmetry if it does not have particles in every representation of the gauge group.  
In other words, the principle of no global symmetries is said to imply the completeness hypothesis.

However, this third bit of lore is false: if $G$ is discrete and non-abelian \cite{Harlow:2018tng} or continuous and disconnected \cite{Heidenreich:2021tna}, the theory need not have any one-form global symmetry, even if its spectrum is incomplete. 
In this paper, we will investigate and refine this bit of lore in the context of  discrete gauge symmetries. 
We will see that although incompleteness of a discrete, non-abelian gauge theory does not necessarily imply the existence of a one-form global symmetry, it does imply the existence of more general topological operators  supported on manifolds of dimension $d-2$.

In discrete $G$ gauge theory,  there  are  certain discrete analogs of the Gukov-Witten  operators \cite{Gukov:2006jk,Gukov:2008sn}, each labeled by a conjugacy class of $G$, that induce a nontrivial holonomy for the Wilson lines.  
In the pure gauge theory without matter, these Gukov-Witten operators are topological and will be denoted as $T_a(M^{(d-2)})$ with $a$ a conjugacy class of $G$.   
If   $G$ is abelian, all the Gukov-Witten operators are symmetry operators, and they generate an electric one-form global symmetry that is isomorphic to $ G$   \cite{Gaiotto:2014kfa}.\footnote{In three spacetime dimensions, the Gukov-Witten operator is not only labeled by a conjugacy class, but also  a representation of the centralizer group of that conjugacy class (see Section \ref{sec:3d}).}   
When $G$ is non-abelian, however, some or all of the Gukov-Witten operators  may not obey a fusion algebra of a group as in \eqref{introgroup}.  
Rather, the fusion algebra takes the general form
\begin{equation}\label{intronon}
T_a (M^{(d-2)}) \times T_b(M^{(d-2)}) = \sum_c N_{ab}^c \, T_c(M^{(d-2)}),
\end{equation}
for some nonnegative integer coefficients $ N_{ab}^c$. 
In the case when $G$ is abelian, $T_a$ and $T_b$ are symmetry operators, and there is only one term in the sum on the right-hand side with coefficient 1. 
But for more general topological operators such as the Gukov-Witten operators in a non-abelian discrete gauge theory, $ N_{ab}^{c}$ may be nonzero for more than one choice of $c$.\footnote{Topological operators that are not associated with any global symmetry have been discussed extensively in two spacetime dimensions  \cite{Frohlich:2004ef,Frohlich:2006ch,Feiguin:2006ydp,Frohlich:2009gb,Aasen:2016dop,Buican:2017rxc,Bhardwaj:2017xup,Chang:2018iay,Ji:2019ugf,Lin:2019hks,Thorngren:2019iar,Pal:2020wwd}.}

In what follows, we will prove several statements regarding the relationship between topological line operators and completeness in discrete $G$ gauge theory weakly coupled to matter in three spacetime dimensions. 
In particular, we will demonstrate that  if there are objects charged under every representation of $G$, i.e., if the spectrum is complete, then every Gukov-Witten line operator ceases to be topological.
 Conversely, if the spectrum is incomplete, then there will be at least one topological Gukov-Witten line.

But there is another statement of completeness we can make in 3d gauge theories. If the gauge theory has matter $\phi$ transforming in some representation $\alpha$, then a Wilson line of representation $\alpha$ can end on a local excitation of the field $\phi$. In the theory with the matter $\phi$, we therefore say that the Wilson line of representation $\alpha$ is \emph{endable}. The condition that the theory is complete is equivalent, therefore, to the statement that all Wilson lines are endable. This naturally suggests a more general notion of completeness: we say that a quantum field theory in three dimensions is \emph{totally complete} if \emph{every} line operator is endable.\footnote{We thank Jacob McNamara for discussions on this point.} We will show that a totally complete 3d field theory has no topological lines whatsoever, and conversely, a 3d field theory that is not totally complete has at least one topological line.

The relationship between completeness and the existence of topological codimension-2 operators can be generalized to higher dimensions. We will focus on 4d discrete gauge theories, arguing that completeness implies the absence of topological  Gukov-Witten surface operators. Conversely, under certain assumptions, the absence of such topological Gukov-Witten surface operators implies that the theory is complete. However, it is not true that completeness implies the absence of \emph{any} topological   surface operators: we will see that 4d $G=\mathbb{Z}_2 \times \mathbb{Z}_2$ gauge theory has a surface operator which remains topological even if the spectrum is complete.

We conclude therefore that within the context of discrete gauge theories, the ``third bit of lore'' regarding the relationship between completeness and the absence of global symmetries should be replaced with a new statement: completeness is not equivalent to the absence of electric one-form global symmetries, but rather to the absence of a particular set of codimension-2 topological operators. This revision naturally suggests a possible modification of the ``first bit of lore'': if $q$-form global symmetries, which are characterized by topological operators of codimension-$(q+1)$, are not allowed in quantum gravity, then perhaps more general topological operators of codimension-$(q+1)$ are also forbidden? In what follows, we will offer some explanations as to why we expect some sort of ``no topological operators'' statement in quantum gravity, but we will leave a more rigorous argument for future work.

The remainder of this paper is structured as follows. In Section \ref{sec:symmetry}, we review higher-form global symmetries and their associated topological operators. In Section \ref{sec:noninvertible}, we introduce more general topological operators that are not necessarily associated with global symmetry. In Section \ref{sec:3d}, we discuss discrete gauge theories in 3d, presenting our main results on the relation between completeness and topological operators, including the illustrative example of $S_3$ gauge theory. In Section \ref{sec:4d}, we explain how these results may be generalized to gauge theories in 4d. In Section \ref{sec:QG}, we discuss topological operators in the context of quantum gravity. Finally, we conclude in Section \ref{sec:CONC} with a discussion of directions for future research.

\section{Symmetry Operators} \label{sec:symmetry}

In this section we discuss various basic aspects of global symmetries and their associated symmetry operators. 
All these discussions are well-known, and we provide a self-contained and streamlined review as a preparation for later sections.

\subsection{Higher-Form Global Symmetry}

An intrinsic description of the global symmetry is given in terms of its symmetry operators.  
It is independent of the explicit Lagrangian presentation (if it exists) of the underlying system.   
This perspective was emphasized and generalized to higher-form global symmetry in \cite{Gaiotto:2014kfa}, which we review below.

In $d$ spacetime dimensions, a $q$-form  global symmetry transformation is implemented by a \textit{symmetry operator} $U_g(M^{(d-q-1)})$ that has support on a codimension-$(q+1)$ closed manifold $M^{(d-q-1)}$.  
Here $g\in G$ is an element of the group $G$. 
When the manifold $M^{(d-q-1)}$ lies at a fixed time, this is an operator acting on the Hilbert space.  
When the manifold $M^{(d-q-1)}$  extends in time, this is a defect that modifies the quantization of the Hilbert space.\footnote{Since we will mostly be  working in Euclidean signature with no choice of a preferred time direction, we will be somewhat cavalier about this distinction and loosely refer to these operators/defects simply as operators when there is no source of confusion.}  

The charged object of a $q$-form global symmetry is an operator ${V}({\cal C}^{(q)})$ supported on a $q$-dimensional closed manifold ${\cal C}^{(q)}$ in spacetime.  The action of  the symmetry on ${V}$ is given by linking a sphere $U_g( S^{d-q-1})$ with ${\cal C}^{(q)}$:\footnote{Here for simplicity we assume the global symmetry is abelian. This is necessarily the case if $q>0$.}
\begin{align}\label{symaction}
U_g( S^{d-q-1})  \cdot {V}({\cal C}^{(q)}) = g({V})  {V}({\cal C}^{(q)})\,,
\end{align}
where $g({V})$ is the representation of $g$ realized by the charged operator ${V}$. This equation is understood physically via shrinking the $S^{d-q-1}$ to a point on the manifold ${\cal C}^{(q)}$, which removes the operator $U_g$ at the expense of introducing the representation $g({V})$.

Physical observables, including correlation functions, can be dressed with these symmetry operators or defects. 
The basic property of the symmetry operators is that correlation functions are invariant under any small deformation of $M^{(d-q-1)}$. 
Thus, symmetry operators are special cases of  \textit{topological operators}.

The symmetry operators obey the group multiplication rule:
\begin{align}\label{group}
U_{g} (M^{(d-q-1)} ) \times U_{g'} (M^{(d-q-1)} ) = U_{g''} (M^{(d-q-1)} ) 
\end{align}
with $g''=  g g' \in G$.  
In particular, any symmetry operator $U_{g} (M^{(d-q-1)} ) $ has an inverse $U_{g^{-1}}(M^{(d-q-1)})$ such that
\begin{align}\label{invertibility}
U_{g} (M^{(d-q-1)} ) \times  U_{g^{-1}} (M^{(d-q-1)} ) = I\,,
\end{align}
where $I$ is the identity operator. 
In Section \ref{sec:noninvertible}, we will discuss more general topological operators whose inverses do not exist.

When $G = U(1)$ and there is a conserved $(q+1)$-form Noether current $j$, the symmetry operator is simply $U_\theta(M^{(d-q-1)}) = \exp[i \theta \int_{M^{(d-q-1)}} \star j]$, with $\theta\in [0,2\pi)$ the $U(1)$ group element. 
The current conservation equation $d (\star j)=0$ implies that the symmetry operator $U_\theta(M^{(d-q-1)})$ is independent of small deformations of $M^{(d-q-1)}$. 
When $M^{(d-q-1)}$ is taken to be the whole space at a given time, the topological nature of  $U_\theta(M^{(d-q-1)})$ means that the $U(1)$ charge is conserved in time.

\subsection{Operators with Boundary}\label{sec:endable}

So far we have discussed operators supported on closed manifolds.  
What about manifolds with boundary?

Starting from a quantum field theory  with certain global symmetry and charged operators $V({\cal C}^{(q)})$, one may couple the system to additional degrees of freedom to obtain a new system.  
Such a coupling may allow the charged operator $V({\cal C}^{(q)})$ to be defined on manifolds with boundary.  
We will say that $V$ is \textit{endable} if it can be defined on a manifold with boundary with a certain choice of boundary operator.  
The choice of such a boundary operator is far from unique, in general. 

If such a coupling renders $V$ endable, it will also break the topological nature of all the symmetry operators $U_g(M^{(d-q-1)})$ that acted nontrivially on $V$ in the pure gauge theory.  
In other words,  the global symmetry is explicitly broken by the coupling. 
To see this, let us assume the contrary.  
We consider a linking configuration between $U_g(S^{d-q-1})$ and $V({\cal C}^{(q)})$ as in the left-hand side of \eqref{symaction}, but now we suppose that ${\cal C}^{(q)}$ is a manifold with boundary. As a result, we may topologically deform $U_g(S^{d-q-1})$ to unlink it from $V$, which is in contradiction with the nontrivial action of $U_g$ on $V$.  Therefore, $U_g$ is no longer a topological operator when $V$ is endable.

For example, consider the $U(1)$ Maxwell theory without matter in four spacetime dimensions. 
The Maxwell theory has a topological surface operator $U_\theta(M^{(2)})=  \exp[{2i\theta\over g^2 } \int_{M^{(2)}}  \star F]$ associated with the $U(1)$ electric one-form global symmetry \cite{Gaiotto:2014kfa}.  
The charged operator is the Wilson line $V({\cal C}^{(1)}) = \exp[i  \oint_{{\cal C}^{(1)}}  A]$.  
We now couple the $U(1)$ gauge theory to a charge +1 scalar field $\phi$.  
In the new theory, the Wilson line $V$ can end on the (gauge non-invariant) field $\phi$.  
Consequently, the one-form global symmetry is broken by this coupling.

\begin{figure}
\begin{center}
\includegraphics[width=15mm]{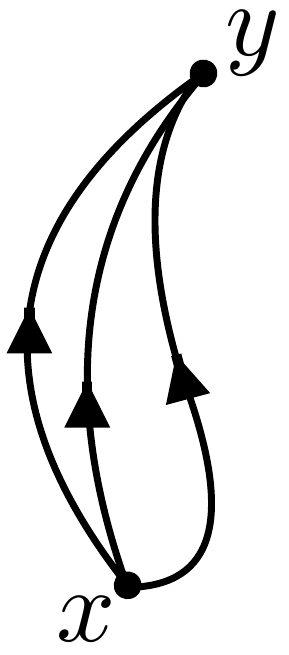}
\end{center}
\caption{Topological line operator $U_g(M^{(1)})$ supported on a manifold with boundary. Physical observables are invariant under deformation of the manifold $M^{(1)}$ stretching between $x$ and $y$ provided the boundary $\partial M = \{ x, y\}$ is held fixed.}
\label{fig:boundary}
\end{figure}

A symmetry operator $U_g(M^{(d-q-1)})$ can also be endable.    
The boundary operator of $U_g(M^{(d-q-1)})$ may or may   not be topological, but the interior of the symmetry operator still is.  
More specifically, all the physical observables are invariant under small deformations of $M^{(d-q-1)}$ while keeping the boundary $\partial M$ fixed (see Figure \ref{fig:boundary}).   
In this sense the  endability of an operator is not in tension with its topological nature.  
When $U_g(M^{(d-q-1)})$ is an endable topological operator, it does not imply that the symmetry is broken.

Let us illustrate the above discussion in the Ising CFT in two spacetime dimensions.  
The Ising CFT has an ordinary $\bZ_2$ global symmetry that is exact and unbroken at the CFT point.  
Its symmetry line $U_g(M^{(1)})$ is endable. 
The endpoint operators are the left- and right-moving free Majorana fermions $\psi,\bar\psi$ and the disorder operator $\mu$,  with conformal weights $(h,\bar h) = (\frac 12, 0 ), (0,\frac 12), ({1\over 16} , {1\over 16} ) $, respectively.\footnote{The endpoint operators are attached to the symmetry line, so they are not local operators of the Ising CFT. It follows that these endpoint operators do not correspond to states in the ordinary Hilbert space under the operator-state correspondence.  Rather, under the operator-state map, they correspond to the states in the twisted Hilbert space where one twists the periodic boundary condition by a $\bZ_2$ action. }  
These endpoint operators are not topological (because of their nontrivial conformal weights), but the interior of the symmetry line is.

There is a more special case in which the boundary operator of an endable topological operator is also topological.  
In this case, the topological operator must act trivially on all operators ${V}(\mathcal{C}^{(q)})$, because we can move the topological boundary to unlink the operator action on the left-hand side \eqref{symaction}.

\subsection{$\bZ_N$ Gauge Theory}\label{ssec:ZN}

Following \cite{Banks:2010zn}, we now illustrate various ideas above using the $\bZ_N$ gauge theory in $d$ spacetime dimensions.  
The $\bZ_N$ gauge theory can be realized using a $U(1)$ one-form gauge field $A^{(1)}$ and a $U(1)$ $(d-2)$-form gauge field $B^{(d-2)}$ with gauge transformation:
\begin{align}
&A^{(1)}\sim A^{(1)} +d\alpha^{(0)}\,,\\
&B^{(d-2)} \sim B^{(d-2)} +d\beta^{(d-3)}\,,
\end{align}
where $\alpha^{(0)}$ and $\beta^{(d-3)}$ are 0- and $(d-3)$-form gauge parameters.  
The Lagrangian for the $\bZ_N$ gauge theory is \cite{Maldacena:2001ss,Banks:2010zn,Kapustin:2014gua,Gaiotto:2014kfa}
\begin{align}\label{BF}
{\cal L}  = {N\over 2\pi } B^{(d-2)} \wedge dA^{(1)}\,.
\end{align}

The $\bZ_N$ gauge theory has a $\bZ_N$ one-form electric global symmetry. 
The symmetry operator is a  topological Wilson operator of the higher form gauge field $B^{(d-2)}$:
\begin{align}
&U _n (M^{(d-2)} ) =  \exp\left[i  n  \oint_{M^{(d-2)}} B^{(d-2)}\right]\,,~~~~~n=0,1,\cdots,N-1\,.
\end{align}
There is also a  $\bZ_N$ $(d-2)$-form magnetic global symmetry. 
The symmetry operator is the topological Wilson line:
\begin{align}
&V _m ({\cal C}^{(1)} ) =  \exp\left[i  m  \oint_{{\cal C}^{(1)}} A^{(1)}\right]\,,~~~~~m=0,1,\cdots,N-1\,.
\end{align}

Note that the symmetry operator of the electric global symmetry is the charged object of the magnetic symmetry, and vice versa.  
Consider an $S^1$ and an $S^{d-2}$ that link with each other. 
We have
\begin{align}\label{ZNlink}
&U_n (S^{d-2}) \cdot V_m (S^1) =  e^{ 2\pi i nm/N} V_m(S^1)\,,\\
&V_m (S^{1}) \cdot U_n (S^{d-2}) =  e^{ 2\pi i nm/N} U_n(S^{d-2})\,.
\end{align}
The first of these equations describes the topological process of shrinking $S^{d-2}$ to a point on $S^1$, whereas the second describes the shrinking of $S^1$ to a point on $S^{d-2}$.
The operator $U _n $ can be interpreted as the Gukov-Witten operator for the gauge field $A^{(1)}$ as it induces a nontrivial holonomy for the Wilson line $V_m$.

In pure $\bZ_N$ gauge theory, none of the above symmetry operators are endable.  
They can only be defined on closed manifolds, and the spectrum of  dynamical charged particles is empty.

We  now introduce additional degrees of freedom to couple the $\bZ_N$ gauge theory to dynamical particles with electric charge $p$  mod $N$ and their composites.  
These dynamical electric particles can be the endpoints of the Wilson lines $V_m$ if $p$ divides  $m$.  
Since $V_m$ with $p|m$ are now endable, $U_n$ with $nm\notin N\bZ$ are no longer topological.   
Hence the electric one-form symmetry is broken from $\bZ_N$ to $\bZ_{gcd(p,N)}$, while the   magnetic $(d-2)$-form symmetry remains $\bZ_N$.  
Note that even though some of the $V_m$'s are endable, they are still all topological.  
More precisely, all physical observables are invariant under small deformations of the interior of the open line ${\cal C}^{(1)}$ that leave its endpoints fixed.  

When $p$ and $N$ are  coprime, the spectrum of electrically charged particles with respect to the ordinary $\bZ_N$ gauge symmetry is complete, and all the Wilson lines are endable. Correspondingly, the Gukov-Witten operators $U_n$ are no longer topological, and the electric one-form symmetry is completely broken.

This establishes the equivalence between the following two statements in the $\bZ_N$ gauge theory:
\begin{itemize}
\item  Completeness of the electrically charged particle spectrum.
\item  Absence of the electric one-form global symmetry,
 \end{itemize}
 or equivalently, between
 \begin{itemize}
\item  Completeness of the electrically charged particle spectrum.
\item  Absence of topological Gukov-Witten operators.
 \end{itemize}
The first version of the equivalence is, however, not true for non-abelian discrete gauge theory.  
By contrast, we will show in Section \ref{sec:3d} that the second equivalence can be generalized to non-abelian discrete gauge group in three spacetime dimensions.  
To address this generalization, we  first need to discuss more general topological operators than the symmetry operators we have considered so far.

\section{Non-Invertible Topological Operators}\label{sec:noninvertible}

Symmetry operators of global symmetry are topological operators.  
In the case when there is a conserved Noether current, the topological nature follows from the local current conservation equation.  
It is natural to ask: is every topological operator associated with a global symmetry?  

The answer is no.  
There are topological operators that are not associated with any global symmetry. 
Such an operator may not obey a group multiplication rule as in \eqref{group}. 
In particular, there might not be an inverse in the sense of \eqref{invertibility}.   
We will refer to such topological operators as the \textit{non-invertible topological operators}.

Consider a non-invertible topological operator  ${T}(M^{(d-q-1)})$ supported on a codimension-$(q+1)$ closed manifold.\footnote{Since the non-invertible operator does not have an inverse, it is not a unitary operator. Hence we will use $T$ to denote it instead of $U$.} 
We would like to define an action of this non-invertible topological operator on the space of $q$-dimensional operators ${V}({\cal C}^{(q)})$.  
The dependence of $V$ on the manifold ${\cal C}^{(q)}$ is generally not topological. 
Similar to \eqref{symaction}, this action is defined by linking a sphere $S^{(d-q-1)}$ with the ${V}({\cal C}^{(q)})$, and then topologically shrinking the sphere to zero size.    We will denote this action by
\begin{align}\label{topaction}
{T}( S^{d-q-1}) \cdot {V}({\cal C}^{(q)}) \,.
\end{align}
Similar to the discussion in Section \ref{sec:endable}, if we deform the system so that $V$ becomes endable, i.e., if it can be defined on a manifold with boundary, then all operators ${T}( S^{d-q-1})$ that act nontrivially on $V$ will cease to be topological.

In two spacetime dimensions, an ordinary global symmetry is implemented by a symmetry line. 
There are many examples of non-invertible topological lines that are not associated with any global symmetry  in two dimensions \cite{Frohlich:2004ef,Frohlich:2006ch,Feiguin:2006ydp,Frohlich:2009gb,Aasen:2016dop,Buican:2017rxc,Bhardwaj:2017xup,Chang:2018iay,Ji:2019ugf,Lin:2019hks,Thorngren:2019iar,Pal:2020wwd}.  
It has been advocated that these non-invertible lines should be viewed as generalizations of the conventional global symmetry \cite{Bhardwaj:2017xup,Chang:2018iay,Lin:2019hks,Thorngren:2019iar}. 
Just like their symmetry line cousins, these non-invertible lines imply strong constraints on renormalization group flows and dualities \cite{Chang:2018iay,nonsymmetry}.

Perhaps the simplest example is the Kramers-Wannier duality line in the Ising conformal field theory (CFT) in two spacetime dimensions  \cite{Frohlich:2004ef,Frohlich:2006ch,Frohlich:2009gb,Aasen:2016dop,Chang:2018iay,Ji:2019ugf,Lin:2019hks}.  
The duality line, denoted as  $T$, obeys the multiplication rule 
\begin{align}\label{isingfusion}
{T}  (M^{(1)})  \times   {T} (M^{(1)})= I  +  U_g(M^{(1)}) \, ,
\end{align}
 where $U_g(M^{(1)})$ is the symmetry line associated with the $\bZ_2$ global symmetry of the Ising CFT, and $g\in \bZ_2$ is the nontrivial element of $\bZ_2$.  
This multiplication rule is not a group multiplication, and indeed the duality line $T$ is not associated with any global symmetry of the Ising CFT.   

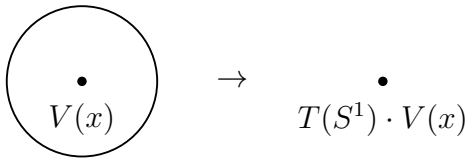
\begin{figure}
\begin{align}\nonumber
\begin{gathered}
\begin{tikzpicture}[scale = 1]
\tikzset{line/.style={line width=0.25mm}}
\draw [line] (0,0) circle (1);
\filldraw [line] (0,0) circle (.05);
\node at (0,-.5) {${V}(x)$};
\node at (2,0) {$\to$};
\filldraw [line] (4,0) circle (.05);
\node at (4,-.5) {${T}(S^1) \cdot {V}(x)$};
\end{tikzpicture}
\end{gathered}
\end{align}
\caption{The duality line $T$ in the two-dimensional Ising CFT defines a (non-invertible) map  on the  local operators  by encircling a local operator ${V}(x)$ and shrinking the circle. This defines an action on the Hilbert space of states via operator-state correspondence.}\label{fig:Naction}
\end{figure}

The duality line acts on the local operators in the Ising CFT as follows:
Let $1$, $\varepsilon$, $\sigma$ be the identity operator, thermal operator, and the spin operator with conformal weights $(h,\bar h)=  (0,0), (\frac 12,\frac 12), ({1\over 16},{1\over 16})$, respectively.  
Then ${T}(S^1)$ acts on them via \eqref{topaction} as (see Figure \ref{fig:Naction}):
\begin{align}
{T}(S^1)\cdot 1 = \sqrt{2} \,,~~~{T}(S^1)\cdot \varepsilon (x)=  -  \sqrt{2} \varepsilon(x)\,,~~~{T}(S^1) \cdot \sigma(x)= 0 \,.
\end{align}

\section{Discrete Gauge Theories in 3d \label{sec:3d}}

\subsection{Topological Line Operators}

We now discuss topological line operators in three spacetime dimensions.  
One-form global symmetries in three dimensions are implemented by symmetry line operators \cite{Gaiotto:2014kfa}.  
In topological quantum field theory (TQFT), these symmetry lines, when extended in the time direction,  are (the probe limits of) the worldlines of  abelian anyons.  
In addition to the symmetry lines, there are also non-invertible topological lines  that are not associated with any global symmetry.\footnote{There are also non-invertible surface operators in three dimension  \cite{Kapustin:2010if}.}   
They are the non-abelian anyons in a TQFT.  
The mathematical framework for describing these lines is the \textit{modular tensor category} (MTC) \cite{Moore:1988qv}.
This is a subject that has been studied extensively in the literature, and here we only give a brief review.

\begin{figure}
\begin{center}
\includegraphics[width=90mm]{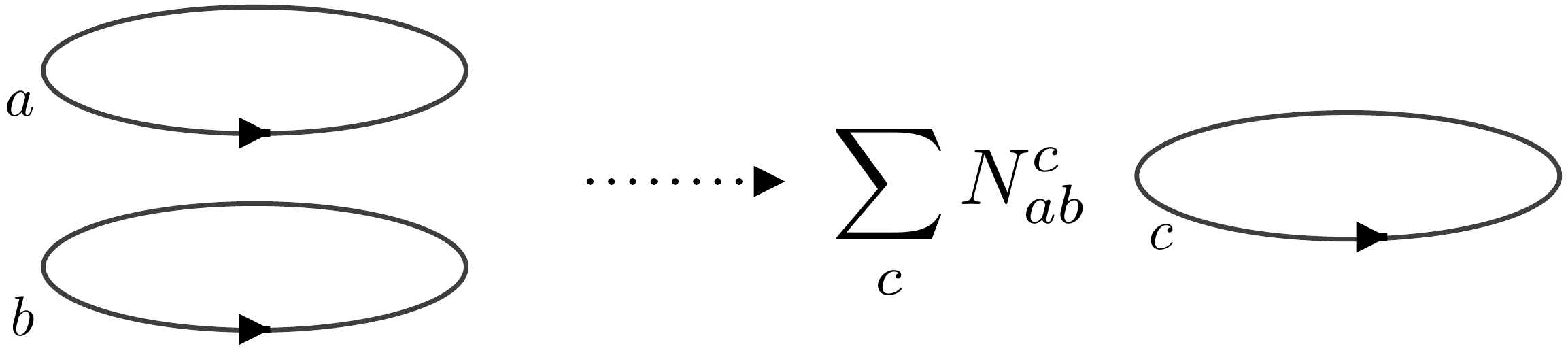}
\end{center}
\caption{Fusion of lines $L_a$ and $L_b$.}
\label{fig:fusion}
\end{figure}

We will denote a topological line  as $L_a$, with $a$ distinguishing the different types of lines.   
In addition to the invertible lines that are associated with the one-form global symmetry, there are also non-invertible topological line operators.  
The fusion of two lines is defined as in Figure \ref{fig:fusion} and takes the form
\begin{equation}
 L_a\times L_b = \sum_{c} N_{ab}^c L_c,
\end{equation}
with $N^c_{ab}$ a non-negative integer.   
The fusion between two lines in three spacetime dimension is commutative, i.e., $L_a\times L_b = L_b\times L_a$.

\begin{figure}
\begin{center}
\includegraphics[width=80mm]{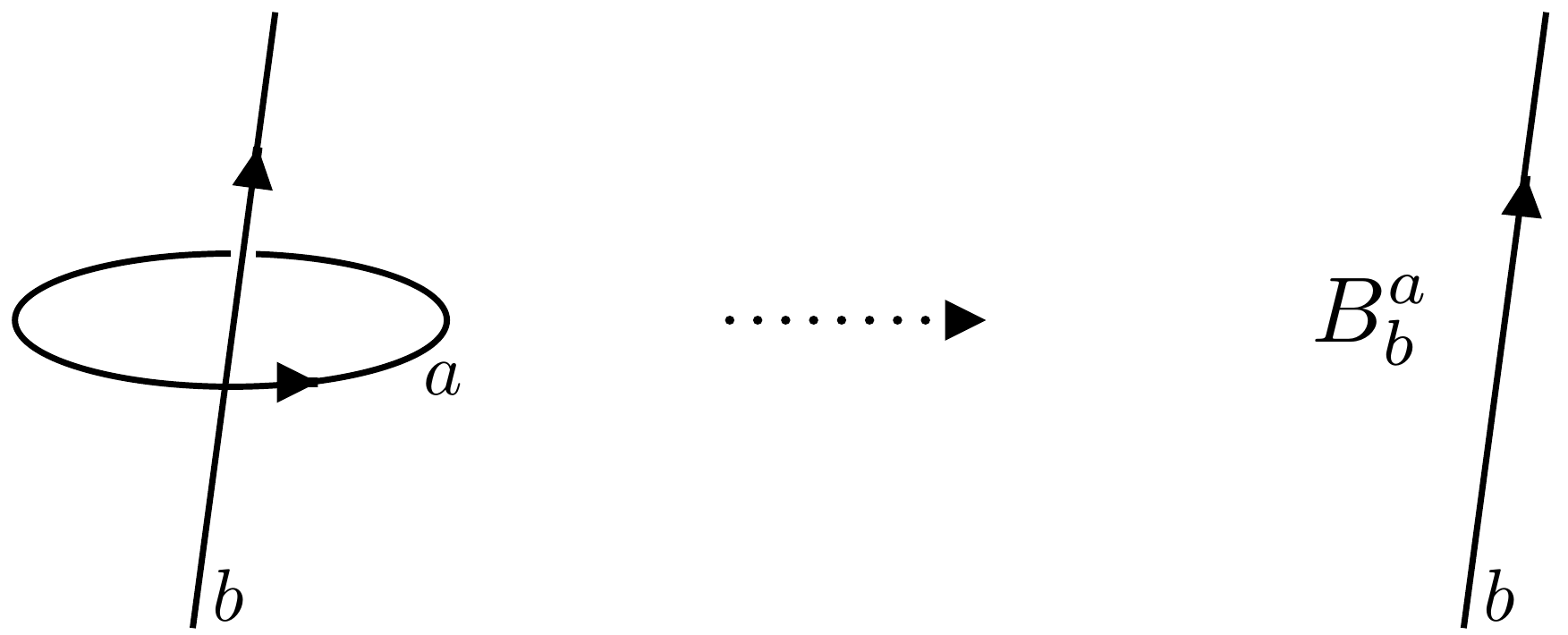}
\end{center}
\caption{We can surround a topological line $L_a$ around another line $L_b$, and then shrink $L_a$ to a point.  This defines an action of $L_a$ on $L_b$ denoted as $L_a \cdot L_b = B^a_b L_b$ with $B^a_b$ the braiding coefficient.  }
\label{fig:braiding}
\end{figure}

As in \eqref{topaction}, we can define an action of the line $L_a$ (which is not necessarily invertible) on another line $L_b$ via braiding.  
More specifically, we surround the line $L_a$ around another line $L_b$ as in Figure \ref{fig:braiding}.  
Since the lines are topological, we can shrink $L_a$ to a point without changing any correlation function,  leaving just the line $L_b$ up to an overall braiding coefficient $B_b^a$:
\begin{align}
L_a \cdot L_b= B_b^a  L_b\,.
\end{align}
Note that generally $L_a \cdot L_b \neq L_b\cdot L_a$.

In a TQFT such as the gauge theory of a finite group with $M$ topological lines, many of the fusion and braiding data are encoded in an $M\times M$ matrix known as the S-matrix.  
The fusion coefficients $N_{ab}^c$ are given in terms of the S-matrix by the Verlinde formula \cite{Verlinde:1988sn}, 
\begin{equation}\label{fusion}
N_{ab}^c = \sum_{n} \frac{ S_{an}  S_{bn}  S_{nc}^*    }{ S_{0n} }.
\end{equation}
The braiding coefficient $B^a_b$ is given by
\begin{equation}
B_b^a = \frac{S_{ab}^*}{S_{0b}^*},
\label{eq:braiding}
\end{equation}
where the index $0$ here represents the trivial line and $*$ stands for complex conjugation. Note that by definition $B_b^0 :=1$, whereas $B_0^a$ may or may not be equal to 1.

\subsection{Discrete Gauge Theories}

In this subsection we review the line operators in the pure gauge theory of a (possibly non-abelian) finite group $G$.  
 We will also discuss the twisted versions of $G$ gauge theory \cite{Dijkgraaf:1989pz}.  
In the context of modular tensor category, such a theory is known as the (twisted) quantum double of $G$ \cite{drinfeld,Kitaev:1997wr}.  
 
In the  pure gauge theory of a finite group $G$, the most familiar class of  topological line operators are the Wilson lines, labeled by a representation of $G$.  
The Wilson line is generally not  an invertible line.  
Relatedly, the fusion rule of the Wilson lines is given by the representation ring of $G$, which is generally not a group.  
More precisely, the invertible Wilson lines are those corresponding to  one-dimensional representations, and they generate the  magnetic one-form symmetry of the $G$ gauge theory.

In addition to the Wilson lines, there are also Gukov-Witten line operators, each of which is labeled by a conjugacy class $[g]$ of $G$.    
These are the discrete analog of the Gukov-Witten operators  in four spacetime dimensions \cite{Gukov:2006jk,Gukov:2008sn}.  
We will denote these lines as $[g]_{\mathbf{1}}$ where $\mathbf{1}$ stands for the trivial representation. 
These lines $[g]_{\mathbf{1}}$  are generally not invertible. 
The invertible ones generate the electric one-form symmetry.

A Gukov-Witten line may also carry electric charge, which corresponds to a nontrivial representation $\alpha$ of the centralizer group $H_{g}$ of an element $g$ in the conjugacy class $[g]$. This possibility arises in three spacetime dimensions because both the Wilson lines and the Gukov-Witten operators happen to be one-dimensional and can be fused with each other.  
Thus, we may denote a general topological line in the $G$ gauge theory as $[g]_\alpha$, with $[g]$ a conjugacy class and $\alpha$ a representation of the centralizer, $H_{g}$. 
Wilson lines $[1]_\alpha$ are the special cases in which the conjugacy class is trivial.  
We will refer to any line $[g]_\alpha$ with a nontrivial conjugacy class as a Gukov-Witten operator, regardless of whether its associated representation $\alpha$ is trivial or nontrivial.

The braiding coefficient $B_b^a$ in (\ref{eq:braiding}) from $L_a= [g]_\alpha$ surrounding a Wilson line $L_b=[1]_\beta$ is given by 
\begin{equation}
B_b^a = \chi_\beta([g])\cdot \text{size}([g]) \cdot { \text{dim}(\alpha) \over \text{dim}(\beta)},
\label{eq:Wilsonbraiding1}
\end{equation}
where $\chi_\beta([g]):= \text{Tr}(\beta(g))$ is the character of an element of the conjugacy class $[g]$ associated with the representation $\beta$, and
 dim$(\alpha)$ is the dimension of the representation $\alpha$. 
This can be derived from the explicit S-matrix of the discrete $G$ gauge theory.  
See, for example,  \cite{1995, Hu_2013}.  
In particular, if $L_b$ is the trivial line $[1]_{\bf 1}$, we have
\begin{equation}
B_0^a =  \text{size}([g]) \cdot \text{dim}(\alpha).
\label{eq:quantumdimension}
\end{equation}
The number $B_0^a$ is known as the \textit{quantum dimension} of the line $L_a$.  
The line $L_a$ is invertible (i.e., it generates a one-form symmetry) if and only if $B_0^a =1$. This means that $[g]_\alpha$ is a symmetry line if only if $\text{size}([g]) =1$ and $ \text{dim}(\alpha) = 1$. The former condition implies that $h g h^{-1} = g$ for all $h \in G$, i.e., $g$ is an element of the center of $G$, and the centralizer of $H_{g}$ is equal to $G$.
From this, we recover the familiar statement that the electric one-form symmetry of $G$ gauge theory, which is generated by the lines $[g]_{\mathbf{1}}$, is given simply by the center of $G$.

In this work, we are interested in discrete $G$ gauge theory coupled to additional matter fields, or particles.  
In the presence of a particle in the representation $\alpha$ of $G$, the Wilson line labeled by $\alpha$ is endable with the endpoint being the matter field.  
Assuming the existence of particles in two representations, $\alpha$, $\beta$ of the gauge group $G$, we may find particles (or multi-particle states, more generally) in additional irreducible representations of the gauge group by considering their tensor product $\alpha \otimes \beta= \sum_\gamma N_{\alpha\beta}^\gamma \gamma.$  
Physically, this corresponds to taking the OPE of the matter fields localized at the endpoints of the Wilson lines in the representations $\alpha$, $\beta$.  
Note that the trivial Wilson line $[1]_{\mathbf{1}}$ is endable, with endpoint being the identity local operator.

\begin{figure}
\begin{center}
\includegraphics[width=40mm]{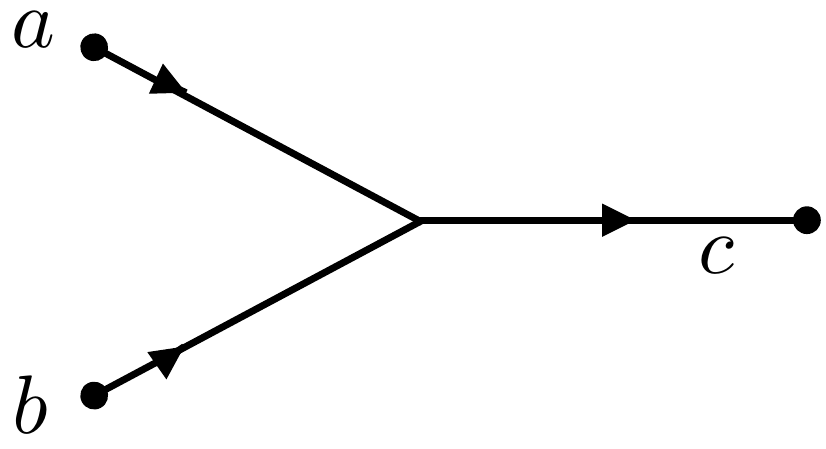}
\end{center}
\caption{Fusion of endable lines $L_a$ and $L_b$. If $L_a$ and $L_b$ are both endable, then any line $L_c$ with $N_{ab}^c \neq 0$ must also be endable. }
\label{fig:oldfusion}
\end{figure}

More generally, we may consider the setup shown in Figure \ref{fig:oldfusion}, in which a pair of lines fuse to form a third line. If the first two lines are endable, their fusion products must be endable as well. 
This means that in a given theory, the set of endable lines are closed under the fusion product \eqref{fusion}.  
To use a more technical term, they form a \emph{fusion subcategory} $\Kend$ of the full category of line operators, $\cC$.  

\begin{figure}
\begin{center}
\includegraphics[width=85mm]{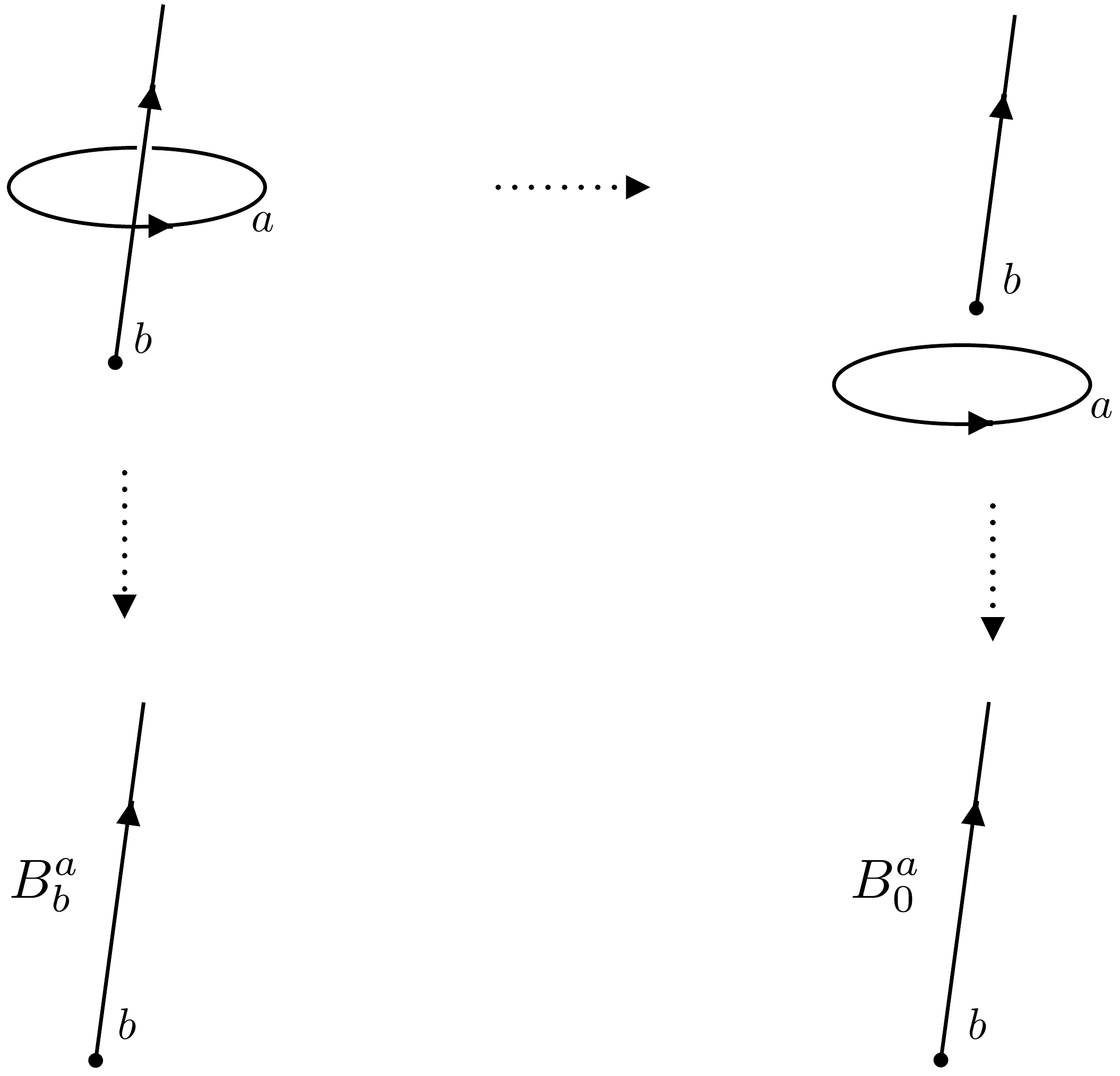}
\end{center}
\caption{Topological and endable lines. If $L_b$ is endable, then $L_a$ may be either shrunk to a point on the line $L_b$ (bottom left), introducing a braiding factor $B_b^a$, or it may be slid off the line (top right) and then shrunk to a point (bottom right), introducing a braiding factor $B_0^a$. If $L_a$ is topological, these two shrinking processes must agree, so we must have $B_b^a = B_0^a$.  }
\label{fig:endable}
\end{figure}

As discussed in Section \ref{sec:endable}, when some lines are endable, other lines might cease to be topological. Consider the setup shown in Figure \ref{fig:endable}, which illustrates the action $L_a \cdot L_b$. If the line $L_b$ can end, the loop $L_a$ may be slid down below the endpoint of $L_b$ and shrunk to a point, giving $L_b$ times a factor $B_0^a$ associated with the shrinking of $L_a$. Alternatively, $L_a$ may be shrunk directly on the line $L_b$, giving $L_b$ times a factor $B_b^a$. Thus, if the line $L_a$ is topological, we must have
\begin{equation}
B_{b}^a = \frac{S_{ab}^*}{S_{0b}^*} = \frac{S_{a0}^*}{S_{00}^*} = B_0^a.
\label{eq:generalsurvive}
\end{equation}
If this condition is not satisfied for all lines $L_b \in \Kend$, the fusion subcategory of endable lines, we conclude that $L_a$ is not a topological line in the theory. Said differently, the lines that remain topological in the presence of charged matter in $\Kend$ are precisely those which braid trivially with the lines in $\Kend$. The lines that braid trivially with all the lines in $\Kend$ form a fusion subcategory of $\cC$, known as the \emph{centralizer} of $\Kend$ in $\cC$, which we denote $C_{\cC}(\Kend)$.

For example, when some Wilson lines are endable, certain Gukov-Witten lines will cease to be topological. 
If the Wilson line $L_b= [1]_\beta$ is endable, then  the lines $L_a= [g]_\alpha$ that remain topological must satisfy (\ref{eq:generalsurvive}):
\begin{equation}
 \chi_\beta([g]) = \chi_\beta([1])\,. 
 \label{eq:topcond}
\end{equation}
In other words, the Gukov-Witten lines that remain topological in the presence of electric matter in representations $\{ \beta \}$ will be those in conjugacy classes $[g]$ whose characters for each of the representations $\{ \beta \}$ are equal to the dimension of the representation $\beta$.

Next, suppose instead that the line $L_a=[g]_\alpha$ is surrounded by the Wilson line $L_b=[1]_\beta$. 
This corresponds to the action $L_b\cdot L_a$. 
In this case, the Wilson line can be shrunk to a point, and the braiding coefficient $B_a^b$ in (\ref{eq:braiding}) is given simply by
\begin{equation}
B_a^b = \chi_{\beta}([g]).
\label{eq:topcond2}
\end{equation}
This means that if the Gukov-Witten line $L_a$ can end, the Wilson line $L_b$ will remain topological only if $\chi_\beta([g]) = \chi_\beta([1]):= \text{dim}(\beta)$. Note that this is the same condition we saw above in (\ref{eq:topcond}). The set of topological Wilson lines will be the ones that satisfy this condition for all endable Gukov-Witten lines $L_a$.

One may also perform a Dijkgraaf-Witten twist \cite{Dijkgraaf:1989pz} on the gauge theory by an element $\omega$ of $H^3(G, U(1))$.  
In a twisted gauge theory, a line is labeled by a conjugacy class $[g]$ as well as a projective representation $\alpha$ of the centralizer group $H_{g}$ of $g$, where the projectivity is determined by $\omega$ and $[g]$. 
Since the conjugacy class $[g]$ for a Wilson line is trivial, the latter is again labeled by a  linear representations of $G$. 
The spectrum of Wilson lines is thus unaffected by the twist, and furthermore the braiding $B_b^a$ of a Wilson line $L_b = [1]_\beta$ and a line $L_a = [g]_{\alpha}$ is still given by (\ref{eq:Wilsonbraiding1})  \cite{1995, Hu_2013}.

\subsection{Completeness and Topological Line Operators}\label{ssec:Completeness}

We now turn to our primary question of interest: the relationship between completeness of spectrum and the existence of topological line operators. There are several different notions of completeness of interest to us. The simplest one, which we refer to  as \emph{completeness}, is the statement that all Wilson lines can end, 
\begin{align}
\text{completeness}:~~ \cW \subset \Kend,
 \end{align}
 where $\cW$ is the set of all Wilson lines.  This follows as a consequence of having matter charged under every representation of the gauge group. For both gauge theories and more general TQFTs in three dimensions, however, we may define a more general notion of completeness: we say that such a theory has a \emph{totally complete} spectrum if every line operator in the theory is endable: $\Kend = \cC$. 

Our main results regarding the relationship between completeness and topological lines in 3d gauge theory with finite gauge group $G$ (untwisted or twisted) are as follows:
\begin{enumerate}
\item The spectrum is complete if and only if there are no topological Gukov-Witten lines.\footnote{In quantum field theory and in gravity, the notion of the gauge group is generally not invariant under duality.  
The distinction between the Wilson lines and the Gukov-Witten lines is also not invariant. 
In this statement, we  have chosen a fixed duality frame in which the theory is naturally described in terms of a discrete gauge theory weakly coupled to electrically charged matter.  }
\item The spectrum is totally complete if and only if there are no topological lines.
\end{enumerate}

The first statement is to be compared with the common belief that the spectrum of a gauge theory is complete if and only if there is no electric one-form global symmetry (see Section \ref{ssec:ZN}).  
This common belief, however, is not true in general. 
For example, the $S_3$ gauge theory in three spacetime dimensions has no invertible Gukov-Witten line, so there is no electric one-form global symmetry regardless of the spectrum.   
A similar example was pointed out in \cite{Harlow:2018tng}.  
Here we propose a natural refinement of the (false) common lore in terms of topological operators.

Some of these results follow from elementary theorems in representation theory of finite groups. 
Others follow directly from a theorem in MTC.  
In this subsection, we will derive these results by a combination of these two approaches.

\subsubsection{Representation Theory Analysis}\label{ssec:Representation}

We start with the first statement on completeness.  

Suppose that $\cW \subset \Kend$, i.e., the spectrum is complete. By (\ref{eq:topcond}), any topological Gukov-Witten line $[g]_\alpha$ must satisfy $\chi_\beta([g]) = \chi_\beta([1])$ for all representations $\beta$ of $G$. But by orthogonality of columns of the character table of $G$, this condition is not satisfied by any $[g] \neq [1]$. 
Thus, completeness $\Rightarrow$ no topological Gukov-Witten lines.

We will prove the converse statement in Section \ref{ssec:MTC}.  
Here we first discuss a weaker version: if $\Kend \subsetneq \cW$, then there must exist at least one topological Gukov-Witten line.   
Note that this is not quite the same as the statement that incompleteness implies the existence of a topological Gukov-Witten line.  
 This is because incompleteness simply requires $\cW \not \subset \Kend$, rather than the stronger condition $\Kend \subsetneq \cW$. Said differently, for the weaker statement we also assume that no Gukov-Witten lines are endable. 

Now we proceed with the proof of the weaker statement.  
Suppose $\cW \subsetneq \Kend$, i.e., the set of endable Wilson lines is a proper subset of the set of endable lines. 
We will show that there must exist  a topological Gukov-Witten line, which  is equivalent to the existence of a nontrivial conjugacy class $[g]$ satisfying $\chi_\beta([g]) = \chi_\beta([1])$ for all $ [1]_\beta \in \Kend$. 
Define the reducible representation
\begin{equation}
\rhoend = \bigoplus_{[1]_\beta \in \Kend} \beta,
\end{equation}
as the direct sum over all irreducible representations corresponding to endable Wilson lines.
Recall that the representations $\{\beta\}$ in $\Kend$ are closed under tensor product of representations, which means that $\rhoend^{\otimes n}$ is necessarily a sum over representations in $\Kend$.
Next, we invoke a theorem of Burnside \cite{Burnside}: if $\alpha$ is a faithful representation of a finite group $G$, then every irreducible representation of $G$ appears in the decomposition of $\alpha^{\otimes n}$ for some $n$.
Since we have assumed $\Kend$ is a proper subset of $\cW$, it follows that $\rhoend$ is not a faithful representation of $G$. 
Thus, there exists some $g \neq 1 \in G$ such that $\rhoend(g)$ is trivial, which by definition of $\rhoend$ implies that $\beta(g)$ is trivial for every $[1]_\beta \in \Kend$. This implies $\chi_\beta([g]) = \chi_\beta([1])$ for all $[1]_\beta \in \Kend$.

\subsubsection{MTC Analysis}\label{ssec:MTC}

We now show that our main statements on  completeness directly follow from a theorem of M\"uger \cite{Muger} in the context of modular tensor category:\footnote{We thank Theo Johnson-Freyd for discussions on this point.}

\vspace{.2cm}
\noindent
\textbf{Theorem} (M\"uger): Let $\mathcal{C}$ be a modular tensor category, and $\mathcal{K}$ a fusion subcategory. Then, $C_{\mathcal{C}}(C_\mathcal{C}(\cK)) = \cK$, and dim$(\cK) \cdot$dim$(C_\cC(\cK)) =$ dim$(\cC)$.

\vspace{.2cm}
\noindent
Here the dimension of a category is defined as the square sum of the quantum dimensions of the lines.  
In other words, M\"uger's theorem establishes a duality pairing between a fusion subcategory $\cK$ and its centralizer $C_\cC(\cK)$.\footnote{See \cite{Hsin:2018vcg} for another physics application of this theorem.}

For our purpose, $\cal C$ is the set of lines in the discrete $G$ gauge theory. 
 We take $\cK = \Kend$ to be the fusion subcategory consisting of all the lines that become endable as we couple to matter fields.  
The centralizer $\centK$ is then the set of lines that remain topological after coupling to matter fields.

If $\Kend = \cC$, i.e., if the spectrum is totally complete, then $C_\cC(\Kend) = \{1 \}$. 
Indeed, $C_\cC({\cal C}) = \{1 \}$ is the  definition of modularity of $\cal C$. 
Physically, this means that every line other than the trivial line will braid nontrivially with some line in $\Kend$, so no nontrivial line remains topological.   
Therefore, total completeness of the spectrum $\Rightarrow$ no topological lines remain.

Conversely, if $\Kend \subsetneq \cC$, then M\"uger's theorem implies $\centKend \neq \{ 1 \}$. This means that if the spectrum is not totally complete, then there will exist nontrivial topological lines that braid trivially with all endable lines, and therefore remain topological in the presence of the charged matter. Thus, the spectrum is totally complete $\Leftrightarrow$ there do not exist any nontrivial topological lines.

This result applies to any TQFT in three spacetime dimensions, regardless of whether or not the theory under consideration is a finite gauge theory. 
Specializing to a finite gauge theory of $G$ (with or without a twist), even more can be said. To begin, recall that for a Wilson line $[1]_\beta$ in a representation $\beta$ of the gauge group $G$, the lines that braid trivially with $[1]_\beta$ are precisely those labeled by conjugacy classes $[g]$ satisfying $\chi_\beta([g]) =  \chi_\beta([1])$. Clearly, this is true for $[g] =[1]$, the trivial conjugacy class, so Wilson lines braid trivially with one another.   
If $\cW$ is the set of Wilson lines, then for any fusion subcategory $\mathcal{W}' \subset {\cal W}$, we have ${\cal W} \subset C_{\cal C} ({\cal W}')$. 

We are now ready to prove that the converse of our first statement on completeness. 
Suppose that there are no topological Gukov-Witten lines, i.e., $\centKend \subset \cW$. Setting $\mathcal{W}' = \centKend$ and using $C_\cC(\centKend) = \Kend$, we learn that $\cW \subset  \Kend$: the spectrum is complete. Thus, completeness $\Leftarrow$ no topological Gukov-Witten lines.   
Combining this with the proof in Section  \ref{ssec:Representation}, we have proven completeness $\Leftrightarrow$ no topological Gukov-Witten lines.

\subsection{Example: $S_3$ Gauge Theory}\label{ssec:S33d}

We now demonstrate the general ideas developed in the previous section in the simplest gauge theory of a non-abelian finite group, the $S_3$ gauge theory.  See  \cite{Cui_2015, Dijkgraaf:1989hb} for a detailed analysis of this theory.

The symmetric group $S_3$ is a group with six elements. The elements of this group are identified with permutations of the set $\{1, 2, 3 \}$ and may be written in cycle decomposition notation. For instance, (123) represents the group element that takes $1 \rightarrow 2$, $2 \rightarrow 3$, $3 \rightarrow 1$, whereas (23) exchanges 2 $\leftrightarrow$ 3.

The conjugacy classes of $S_3$ are labeled by partitions of $3$, namely $[3]$, $[2,1]$, and $[1^3]$, corresponding to cycles of size 3, 2, and 1 (the identity element), respectively. The number of elements in each of these conjugacy classes, i.e., the size of the conjugacy class, is 2, 3, and 1, respectively. An element of the conjugacy class $[3]$, such as $(123)$, has a $\mathbb{Z}_3$ centralizer subgroup: namely, $(123)$ is stabilized by the identity, $(123)$, and $(132)$. An element of the conjugacy class $[2,1]$, such as $(23)$, has a $\mathbb{Z}_2$ centralizer subgroup: namely, $(23)$ is stabilized by the identity and $(23)$. 
Since $H^3(S_3, U(1))$ is trivial,  the $S_3$ gauge theory does not admit a Dijkgraaf-Witten twist.

In three spacetime dimensions, pure $S_3$ gauge theory includes eight topological line operators. Three of these are the Wilson lines, which  correspond to the trivial conjugacy class $[1^3]$ of $S_3$ and are labeled by representations of $S_3$. We denote these line operators by $1$, $1_-$, and $1_{\mathbf{2}}$, corresponding to the trivial representation, the sign representation, and the standard, 2-dimensional representation of $S_3$, respectively. Next, there are two line operators corresponding to the conjugacy class $[2,1]$, labeled by the two representations of the centralizer $\mathbb{Z}_2$. We denote these line operators by $\tau_+$, $\tau_-$, respectively. Finally, there are three line operators corresponding to the conjugacy class $[3]$, labeled by the three representations of the centralizer $\mathbb{Z}_3$. We denote these line operators by $\theta_0$, $\theta_1$, and $\theta_2$. 

The quantum dimensions of these operators are dim$(1) =$ dim$(1_-) = 1$, dim$(1_{\mathbf{2}}) = $ dim$(\theta_\mu) = 2$, and dim$(\tau_\pm) = 3$.  
(Here $\mu=0,1,2$.) 
Only the trivial line and the Wilson line $1_-$ are symmetry operators.  
They generate a $\bZ_2$   magnetic one-form global symmetry.  
The other six lines are all non-invertible.  
In particular, since all the Gukov-Witten operators are non-invertible,  there is no electric one-form global symmetry.

\begin{table}
\centering
\begin{tabular}{|c|c|c|c|c|c|c|c|c|c|}
\hline
  \multicolumn{2}{|c|}{\multirow{2}{*}{}}&  \multicolumn{3}{|c|}{$[1^3]$} & \multicolumn{2}{|c|}{[2,1]} & \multicolumn{3}{|c|}{[3]}   \\ \cline{3-10}
 \multicolumn{2}{|c|}{} &  1& $1_-$ &$1_{\mathbf{2}}$ &$\tau_+$ & $\tau_-$ & $\theta_0$ & $\theta_1$& $\theta_2$ \\ \hline
 \multirow{3}{*}{$[1^3]$} &1 &1 &1 &2 &3 &3 &2 &2 &2 \\ \cline{2-10}
 &$1_-$ & 1& 1& 2& -3& -3& 2& 2& 2 \\ \cline{2-10}
& $1_{\mathbf{2}}$ & 2& 2& 4& 0& 0& -2& -2& -2 \\ \hline
 \multirow{2}{*}{$[2,1]$} &$\tau_+$ &3& -3& 0& 3& -3& 0& 0& 0 \\ \cline{2-10}
& $\tau_-$ & 3& -3& 0& -3& 3& 0& 0& 0 \\ \hline
 \multirow{3}{*}{$[3]$} &$\theta_0$ & 2& 2& -2& 0& 0& 4& -2& -2 \\ \cline{2-10}
 &$\theta_1$ & 2& 2& -2 & 0& 0& -2& -2& 4 \\ \cline{2-10}
&$\theta_2$ & 2& 2& -2& 0& 0& -2& 4& -2 \\ \hline
\end{tabular}
\caption{The S-matrix for the eight topological lines in $S_3$ gauge theory (multiplied by an overall factor of $6$).}
\label{tab:S3S}
\end{table}

The S-matrix for the eight line operators is shown in Table \ref{tab:S3S}.  The fusion rules between these line operators are:
\begin{align}
1_- \times 1_- = 1 \,,~~~ 1_- \times 1_{\mathbf{2}} = 1_{\mathbf{2}} \,,~~~ & 1_{\mathbf{2}} \times 1_{\mathbf{2}} = 1 + 1_- + 1_{\mathbf{2}}  \nonumber\\
 \theta_\mu \times 1_- = \theta_\mu \,,~~~~ \theta_\mu \times \theta_\mu = 1 + 1_- + \theta_\mu \,,~~~ & \theta_\mu \times \theta_{\nu \neq \mu} = 1_{\mathbf{2}} + \theta_{\lambda\neq \mu, \nu}\,,~~~  \theta_\mu \times 1_{\mathbf{2}} = \sum_{\nu \neq \mu} \theta_\nu  \nonumber\\
 1_- \times \tau_{\pm} = \tau_{\mp} \,,~~~ \tau_+ \times \tau_+ = \tau_- \times \tau_- = 1 + 1_{\mathbf{2}} &+  \sum_{\mu} \theta_\mu\,,~~~ \tau_+ \times \tau_- = 1_- + 1_{\mathbf{2}} +  \sum_{\mu} \theta_\mu\nonumber \\
  1_{\mathbf{2}} \times \tau_{\pm} =  \theta_{\mu} &\times \tau_{\pm} = \tau_+ + \tau_-.
\end{align}
We may also read off the braiding coefficients $B_b^a$ from the S-matrix using (\ref{eq:braiding}). Using Table \ref{tab:S3character}, one can easily check that for $L_a = [g]_\alpha$ a Gukov-Witten line and $L_b = [1]_\beta$ a Wilson line, the braiding coefficients defined as \eqref{eq:braiding} obey (\ref{eq:Wilsonbraiding1}).\footnote{For instance, for $L_a = \theta_0$ and $L_b = 1_{\mathbf{2}}$, we have 
$B_{b}^a = \chi_{\bf 2}([3])\cdot \text{size}([3]) \cdot { \text{dim}( { 1}) \over \text{dim}(\mathbf{2})} = - 1 \cdot 2 \cdot \frac{1}{2} = - 1$,
which is the same as $\frac{S_{ab}^*}{S_{0b}^*} = \frac{-2}{2} = -1$.
}

From the fusion rules, we can determine the allowed possibilities for the spectrum of charged matter, or equivalently, the consistent choices of fusion subalgebra $\Kend$.\footnote{An inconsistent spectrum, for instance, would be one with only a  particle in the  representation ${\mathbf{2}}$ of $S_3$. By consider the fusion of two such particles, we obtain a particle in the sign representation   of $S_3$.  Hence any consistent spectrum containing the former particle necessarily contains the latter particle as well.} There are eight possible choices for the allowed spectrum of matter $\Kend$:
\begin{enumerate}[(i)]
\item 1
\item 1, $1_-$
\item 1, $1_-$, $1_{\mathbf{2}}$ (complete)
\item 1, $1_-$, $\theta_0$
\item 1, $1_-$, $\theta_1$
\item 1, $1_-$, $\theta_2$
\item 1, $1_-$, $1_{\mathbf{2}}$, $\theta_0$, $\theta_1$, $\theta_2$ (complete)
\item 1, $1_-$, $1_{\mathbf{2}}$, $\theta_0$, $\theta_1$, $\theta_2$, $\tau_+$, $\tau_-$ (totally complete)
\end{enumerate}

\begin{table}
\centering
\begin{tabular}{|c|c|c|c|} \hline
Rep./Conj. class & $[1^3]$ & $[2,1]$ & $[3]$ \\ \hline
Trivial & 1 &1 & 1 \\ \hline
Sign &  1 & -1 & 1 \\ \hline
Standard & 2 & 0 & -1 \\ \hline
\end{tabular}
\caption{The character table for the group $S_3$.}
\label{tab:S3character}
\end{table}

According to the theorem of M\"uger discussed above, the centralizer $C_\cC(\Kend)$ of one of the above fusion subcategories $\Kend$ is itself a fusion subcategory, with $C_\cC(C_\cC(\Kend)) = \Kend$. 
The category $\Kend$ and its centralizer $C_\cC(\Kend)$ form a dual pair in the above sense.  
Using the S-matrix in Table \ref{tab:S3S}, we find the following dual pairs:
\begin{equation}
\cK_{\rm (i)} \leftrightarrow \cK_{\rm (viii)}\,,~~~\cK_{\rm (ii)} \leftrightarrow \cK_{\rm (vii)}\,,~~~\cK_{\rm (iii)} \leftrightarrow \cK_{\rm (iii)}\,,~~~\cK_{\rm (iv)} \leftrightarrow \cK_{\rm (iv)}\,,~~~\cK_{\rm (v)} \leftrightarrow \cK_{\rm (vi)}.
\label{eq:S3dual}
\end{equation}

If we couple the $S_3$ gauge theory to  matter fields  in some $\cK_{\alpha}$, then the corresponding lines in $\cK_{\alpha}$ can end, and the lines that remain topological are precisely those in $C_\cC(\cK_\alpha)$. 
For instance, if the spectrum is complete but has no magnetic matter, then all Wilson lines are endable, so $\Kend = \cK_{\rm (iii)}$.  
The topological lines are those in $C_\cC(\cK_{\rm (iii)}) = \cK_{\rm (iii)}$, i.e., the topological lines will be simply the Wilson lines themselves. 
 On the other hand, a spectrum that is totally complete will have matter in $\cK_{\rm (viii)}$, and since $C_\cC(\cK_{\rm (viii)})=\cK_{\rm (i)} = \{1\}$, only the trivial line will remain topological. From the dualities in (\ref{eq:S3dual}), we see that as expected, nontrivial topological lines exist if and only if the spectrum is not totally complete.

Theories with $\Kend = \cK_{\rm (iii)}$, $\cK_{\rm (vii)}$, or $\cK_{\rm (viii)}$ are complete. The duals to these three subcategories are given respectively by $\centKend = \cK_{\rm (iii)}$, $\cK_{\rm (ii)}$, and $\cK_{\rm (i)}$, which are precisely the three fusion subcategories without any Gukov-Witten lines. We see that topological Gukov-Witten lines exist if and only if the spectrum is incomplete.

\section{Discrete Gauge Theories in 4d \label{sec:4d}}

Many aspects of our 3d analysis carry over to finite gauge theories in 4d, but there are also some important differences. In particular, while there are still Wilson line operators labeled by representations $\alpha$ of the gauge group $G$, the Gukov-Witten operators are now \emph{surface} operators, labeled by conjugacy classes $[g]$ of $G$, which live on 2-dimensional manifolds in spacetime. The latter are the discrete version of the Gukov-Witten operators \cite{Gukov:2006jk,Gukov:2008sn}. 
Whereas the endpoints of Wilson lines are associated with the creation of electrically-charged particles, the boundaries of Gukov-Witten surfaces are associated with the creation of  magnetically-charged strings. 
Lines can fuse to give other lines, and  surfaces can fuse to give other  surfaces.

A line and a surface can have nontrivial linking in four spacetime dimensions. 
The linking coefficient between a Wilson line  and a Gukov-Witten surface   is the same as in the untwisted $G$ gauge theory in 3d: if a Gukov-Witten surface $T_{[g]}$ of conjugacy class $[g]$ surrounds a Wilson line $L_{\beta}$ of representation $\beta$, the surface can be shrunk to a point on the line at the expense of a linking coefficient,
\begin{equation}
B_\beta^{[g]} =\frac{\chi_\beta([g])}{\chi_\beta([1])} \cdot \text{size}([g])  \,,
\end{equation}
as in equation (\ref{eq:Wilsonbraiding1}). The equivalence of these braiding and linking coefficients in 3d and 4d can be justified via dimensional reduction of the 4d gauge theory \cite{Moradi:2014cfa, Lan:2018vjb,Lan:2018bui}.
 Similar linkings between the Wilson lines and codimension-2 Gukov-Witten operators in discrete $G$ gauge theory exist in higher spacetime dimensions.

\begin{figure}
\begin{center}
\includegraphics[width=35mm]{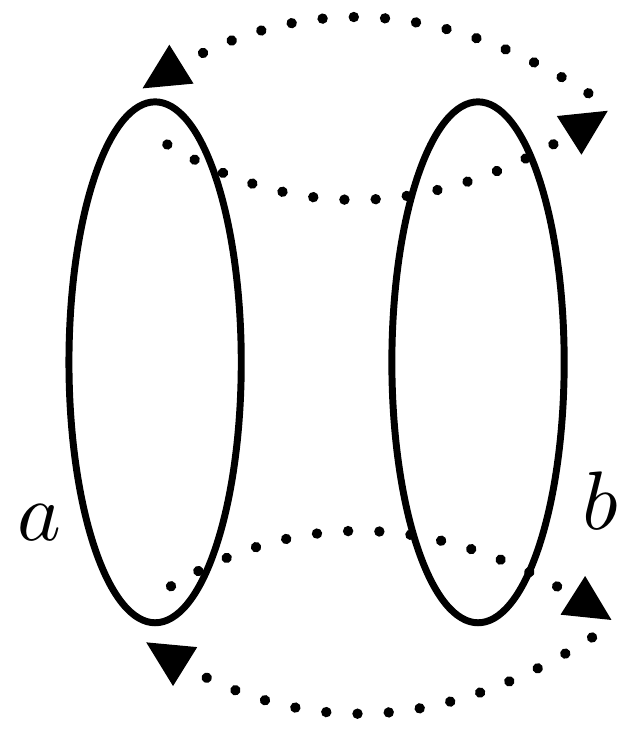}~~~~~~~~~~~~~
\includegraphics[width=65mm]{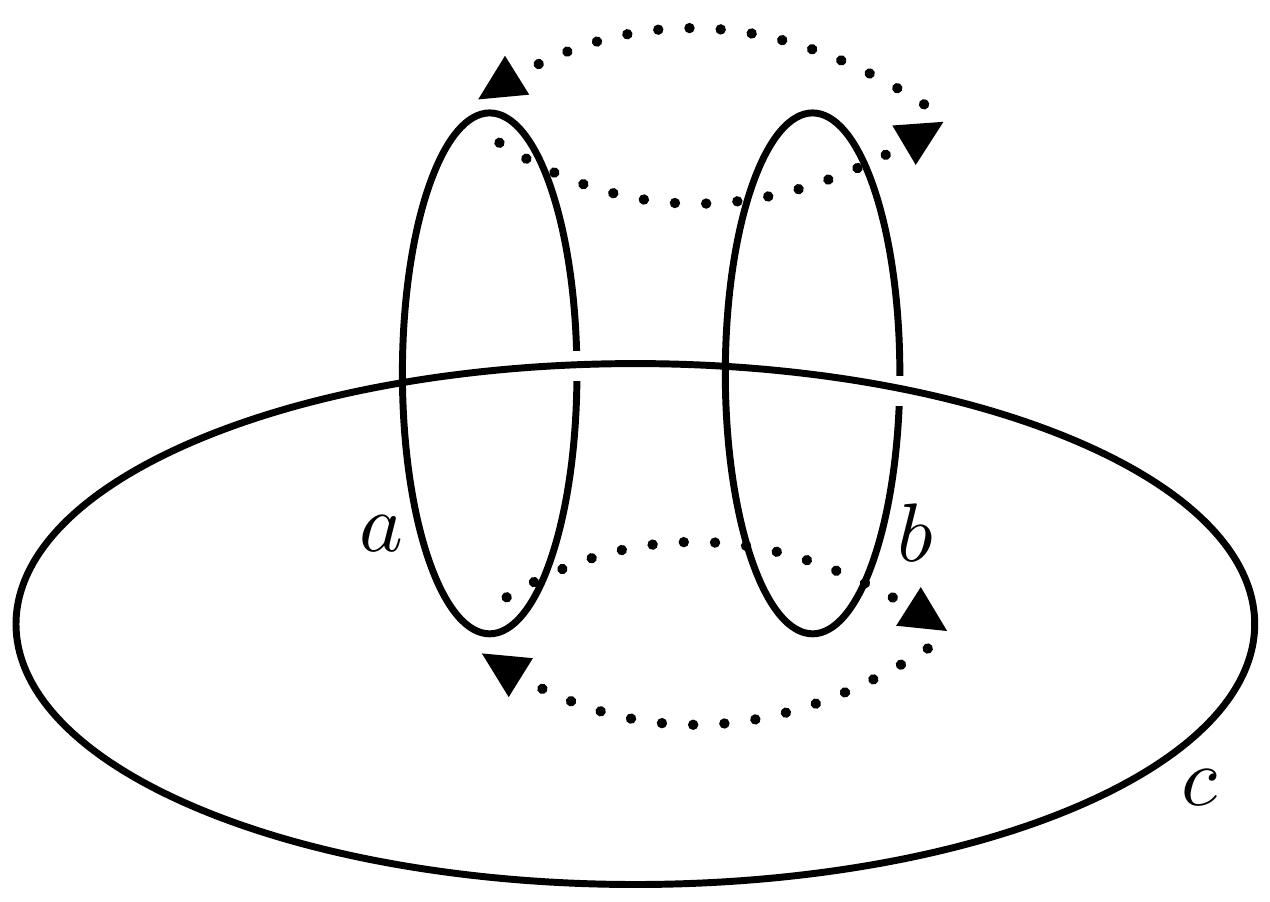}
\end{center}
\caption{Double (left) and triple (right) linking  of surfaces in 4d.  We depict the linking process for  surface defects that extend in time.  At a fixed time, each surface defect is one-dimensional as shown.  The dotted lines describe the motion of the surface defect in the linking process.}
\label{fig:multisurface}
\end{figure}

In addition, two or even three surfaces may link with each other \cite{Wang:2014xba,Jiang:2014ksa} (see also \cite{Wang:2014oya,Wang:2014wka,Putrov:2016qdo,Wang:2018iwz}).  
Figure \ref{fig:multisurface}, adapted from \cite{Wang:2014xba}, shows how these linkings can be visualized for surfaces that extend in time and are one-dimensional loops in space.

In 3d, a line ceases to be topological if it braids nontrivially with a line that can end. In 4d, a similar phenomenon occurs: a line ceases to be topological if it links nontrivially with a surface that can end, and a surface ceases to be topological if it links nontrivially with a line that can end. However, additional possibilities arise in 4d due to the two and three-surface linking: a surface will cease to be topological if it links nontrivially with another surface that can end, and if two surfaces have non-trivial three-surface linking with a third surface that can end, then at least one of those surfaces cannot be topological.

Another new feature in 4d is that a Gukov-Witten surface $T$ can be shrunk to a line \cite{Moradi:2014cfa}. The shrinking function $\cS$: $\{\text{surfaces}\}$ $\rightarrow$ $\{\text{lines}\}$ must commute with fusion, $\cS(T_1) \times \cS(T_2) = \cS(T_1 \times T_2)$. Shrinking represents a smooth deformation of the surface, which means that if $T$ is topological, then $\cS(T)$ must also be topological. Similarly, if $T$ is endable, then $\cS(T)$ must also be endable.

In this paper, we focus our attention on the endability of line operators, which simply involves the addition of charged particles to the spectrum of the theory, so we assume that no surfaces are endable. Under this assumption, we need not worry about two-surface or three-surface linking.  
 A surface $T_{[g]}$   ceases to be topological if and only if it links nontrivially with a line $L_\beta$  that can end, i.e., if $B_\beta^{[g]} \neq B_0^{[g]}$, with 0 stands for the trivial line. Once again, this is equivalent to the condition that
\begin{equation}
 \chi_\beta([g]) \neq  \chi_\beta([1]) := \text{dim}(\beta).
\end{equation}
From our analysis in Section \ref{ssec:Completeness}, we know that if the set of endable lines $\Kend$ is a proper subset of Wilson lines that is closed under fusion, then there exists at least one nontrivial conjugacy class $[g]$ such that $\chi_\beta([g]) =  \chi_\beta([1])$ for all endable Wilson lines $L_\beta \in \Kend$. This means that if the spectrum is incomplete, so that not all Wilson lines can end, then there exists at least one topological Gukov-Witten surface. Conversely, if the spectrum is complete, then there do not exist any topological Gukov-Witten surfaces, as any Gukov-Witten surface, labeled by a conjugacy class $[g]$, links nontrivially with at least one Wilson line. 

However, completeness does not preclude the existence of other topological surfaces in the theory. This may occur, for instance, in the case of $\mathbb{Z}_2 \times \mathbb{Z}_2$ gauge theory, which we now consider in more detail.\footnote{We thank Po-Shen Hsin for pointing this out to us, and to Meng Cheng and Qing-Rui Wang for illuminating discussions.}

\subsection{A One-Form Symmetry that Acts Trivially on All Lines}

Many of the details of $G=\mathbb{Z}^{(1)}_2 \times \mathbb{Z}^{(2)}_2$ gauge theory follow straightforwardly from our previous discussions. Each $\mathbb{Z}_2^{(i)}$ has a Wilson line $e_i$ and a Gukov-Witten surface $m_i$, with nontrivial linking coefficients $B_{e_1}^{m_1} = B_{e_2}^{m_2} = -1$. In addition, the Wilson lines fuse according to $e_1 \times e_2 = e_1 e_2$, and the Gukov-Witten surfaces fuse as $m_1 \times m_2 = m_1 m_2$. Together $e_1$, $e_2$, and $e_1 e_2$ represent the three nontrivial representations of $G$, whereas $m_1, m_2$, and $m_1 m_2$ represent the three nontrivial conjugacy classes.

This theory admits a simple Lagrangian description:
\begin{equation}
\mathcal{L}= \frac{2}{2 \pi} A_1d  B_1 + \frac{2}{2 \pi} A_2 d B_2 ,
\end{equation}
with each $A_i$ a one-form $U(1)$ gauge field and each $B_i$ a two-form $U(1)$ gauge field. 
The Wilson lines and the Gukov-Witten surface operators are respectively given by
\begin{equation}
e_i(\gamma) = \exp\left(i \oint_\gamma A_i \right) \,,~~~m_i(\Sigma) = \exp\left( i \oint_\Sigma B_i\right).
\end{equation}
where $\gamma$ is a closed  curve and $\Sigma$ is a closed two-dimensional surface.  

So far, the story is very similar to the case of $\mathbb{Z}_N$ gauge theory studied in Section \ref{ssec:ZN}. 
However, as discussed in \cite{Hsin:2019fhf} (see also \cite{PhysRevB.96.045136}), there is a surface operator $v(\Sigma)$ 
 in the $\mathbb{Z}_2 \times \mathbb{Z}_2$ gauge theory that is associated with the nontrivial element of $H^2(\bZ_2\times \bZ_2,U(1))=\bZ_2$.  
This surface   generates a $\mathbb{Z}_2$ one-form global symmetry. 
However, this one-form symmetry acts trivially on all the Wilson line operators in the theory. 
Instead, 
the surface $v$ has a nontrivial triple linking with the surfaces $m_1$ and $m_2$:
\begin{equation}
\langle v(\Sigma) m_1(\Sigma') m_2(\Sigma'') \rangle = (-1)^{\text{Tlk}(\Sigma, \Sigma', \Sigma'')},
\label{eq:triplebraiding}
\end{equation}
where Tlk$(\Sigma, \Sigma', \Sigma'')$ is the triple linking number \cite{Carter_2003} between the surfaces $\Sigma$, $\Sigma'$, and $\Sigma''$.

What does the surface $v$ mean for the relationship between completeness and global symmetries? Since $v$ links trivially with all Wilson lines, it remains topological even if the electric spectrum is complete. 
In order to destroy the topological nature of the $v$ surface, one would need to add not just electrically charged particles, but also magnetically charged strings, which allow the surfaces $m_1$ and $m_2$ to end and render $v$ non-topological due to its triple linking with $m_1$ and $m_2$.
Thus, it is not true that completeness of the electrically charged particle spectrum implies  the absence of  any topological surfaces whatsoever in 4d gauge theory with finite gauge group $G$. Instead, completeness implies that a particular set of surfaces are not topological -- namely, the set of Gukov-Witten surfaces, which link nontrivially with the Wilson lines. 
Quite possibly a more general notion of completeness, which demands endability of surfaces as well as lines, may be related to the absence of any topological operators. We leave a more detailed analysis for future work.

\section{Topological Operators and Quantum Gravity \label{sec:QG}}

We have seen that in a gauge theory of a finite (possibly non-abelian) group $G$, an incomplete spectrum does not necessarily imply the existence of an electric one-form global symmetry. 
Instead, there will typically be non-invertible topological operators of codimension-2, which braid non-trivially with the Wilson lines in the theory. 
These observations support  the perspective advocated in \cite{Bhardwaj:2017xup,Chang:2018iay,Lin:2019hks,Thorngren:2019iar} that topological operators  should be viewed as generalization of (higher-form) global symmetries.

Standard lore holds that global symmetries are not allowed in a consistent theory of quantum gravity.  
Topological operators are natural generalizations of global symmetries. 
It is then natural to conjecture that \textit{these topological operators, like symmetry operators, should also be excluded in a consistent theory of quantum gravity.} 
In other words, they reside in the Swampland \cite{Vafa:2005ui}.

Several lines of evidence support this conjecture. Here we present three:
\begin{enumerate}
\item 
\textbf{Summing over topologies}: 
It is difficult to even define a topological operator (which includes  symmetry operators for global symmetries)  in quantum gravity in complete generality. 
In quantum field theory on a fixed geometry, topological operators can be abstractly defined by their correlation functions subject to various consistency conditions, such as crossing and operator product expansion. 
The topological operators are inserted on fixed, nontrivial cycles of the spacetime manifold.  
The correlation function only depends  on the topological class of the support of these topological operators, not on the detailed shape. 
By contrast, in quantum gravity,  we are instructed to sum over all possible spacetime topologies.  
It is therefore meaningless to insert a topological operator on a fixed cycle of the spacetime manifold.  
This makes it difficult to derive any physical observables of these topological operators in quantum gravity.\footnote{We thank Luca Iliesiu for related discussions.}  
See \cite{McNamara:2020uza} for a related discussion on the absence of operators with compact support in quantum gravity. 
\item \textbf{Completeness hypothesis}: Quantum gravity lore holds that the spectrum of a gauge theory must be complete \cite{Polchinski:2003bq, Banks:2010zn, Harlow:2018tng}.  
Given that completeness of the spectrum is tied to the absence of certain topological operators, the completeness hypothesis provides further supporting evidence for the absence of topological operators in quantum gravity.
\item \textbf{Fusion algebra}: The non-invertible  topological operators and the symmetry operators together form a nontrivial fusion ring  \eqref{intronon}.  
Generally, the symmetry operators can appear  in the fusion channel of two non-invertible topological operators. 
For instance, in the 2d Ising CFT, the $\bZ_2$ symmetry operator $U_g$ appears in the fusion of a pair of non-invertible topological $T$ operators (see \eqref{isingfusion}).  
Hence if the symmetry operators are forbidden in quantum gravity, so are the topological operators whose fusion products involve the symmetry operators.\footnote{There are, however, topological operators whose fusion products do not involve any symmetry operators, in which case this argument does not hold.  For example, there is a non-invertible topological line $W$ in the 2d tricritical Ising CFT obeying the fusion algebra $W\times W=I+W$ \cite{Feiguin:2006ydp,Chang:2018iay}.}
\end{enumerate}

\begin{figure}
\begin{center}
\includegraphics[width=100mm]{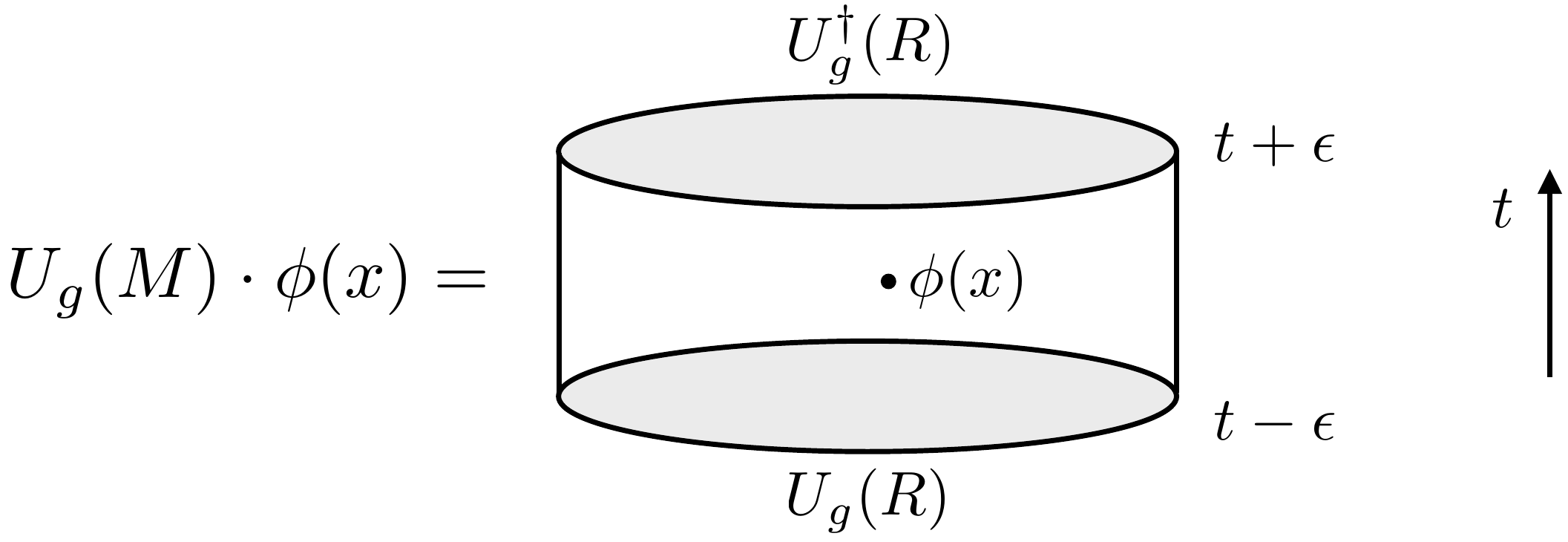}
\end{center}
\caption{The action of a topological defect on a local operator $U_g(M) \cdot \phi(x)$.  
Here $R$ is a disk in space  and the closed manifold $M$, which has topology of a sphere, is  the boundary of $R\times [t-\epsilon ,t+\epsilon]$.}
\label{fig:UM}
\end{figure}

In the context of AdS$_d$/CFT$_{d-1}$, Harlow and Ooguri  \cite{Harlow:2018tng} have given an argument for the absence of $q$-form global symmetries in AdS$_d$ quantum gravity (with $d \geq q+3$) using entanglement wedge reconstruction \cite{Czech:2012bh, Wall:2012uf, Headrick:2014cta, Dong:2016eik}.  
However, we will see that there is a difficulty in extending their argument to exclude more general topological operators. 
For simplicity, we will focus on the case of codimension-1 topological operators.

In Euclidean spacetime, we have defined an action \eqref{symaction} of a symmetry defect $U_g(M)$ on a local operator $\phi(x)$ by encircling the former around the latter, denoted as $U_g(M) \cdot \phi(x)$.   
In the case when the manifold $M$ is shown  in Figure \ref{fig:UM}, this action reduces to:
\begin{align}
U_g(M) \cdot \phi(x)  =  U_g^\dagger(R) \phi(x) U_g(R),
\end{align}
where $U_g(R)$ is a symmetry operator  at a fixed time supported on a disk $R$ in space. 
In the argument of \cite{Harlow:2018tng}, it was crucial that the symmetry operator is \textit{splittable}.   
This implies that if the spatial region $R$ can be written as a disjoint union of subregions $R_i$, i.e.,  $R=\cup_i R_i$, then $U_g(M)$ must obey
\begin{align}\label{splittable}
U_g(M) \sim \prod_i U_g (M_i),
\end{align}
where $M_i$ is a closed manifold associated with   $R_i$ defined similarly as in  Figure \ref{fig:UM}.  
Here, the symbol $\sim$ means that the equality is true up to ($c$-number) phases from the 't Hooft anomaly of the global symmetry.\footnote{See \cite{Chang:2018iay,Lin:2019kpn,Cordova:2019wpi} for an example of an anomalous phase for a $\bZ_2$ global symmetry in 2d. 
}

\begin{figure}[h!]
\centering
\includegraphics[width=.9\textwidth]{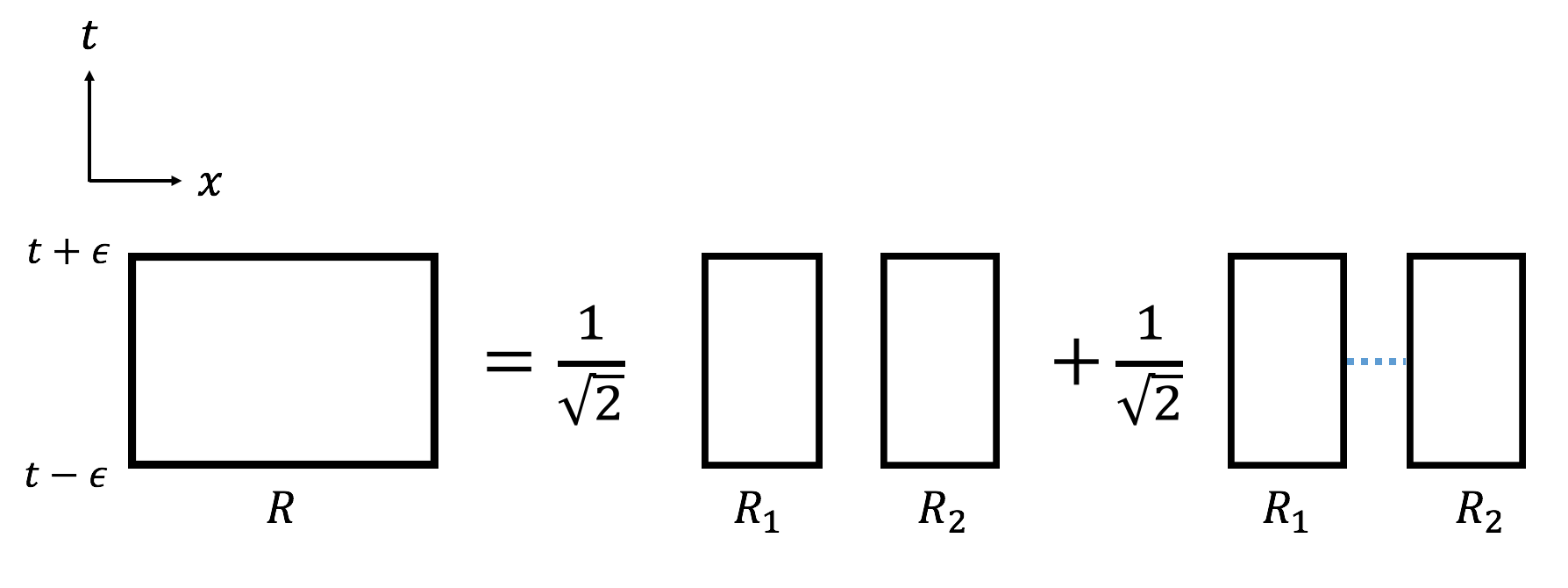}
\caption{The splitting of the  non-invertible topological line $T(M)$ (shown in black on the left) in the 2d Ising CFT, where $M$ is the boundary of the rectangle $R\times [t-\epsilon,t+\epsilon]$.  On the right-hand side of the equality, we have $T(M_1)$ and $T(M_2)$ (shown in black), as well as a $\bZ_2$ symmetry line $U_g$  joining them (shown in the dotted blue line) in the last term.   }\label{fig:nonsplit}
\end{figure}
\begin{figure}[h!]
\centering
\includegraphics[width=.6\textwidth]{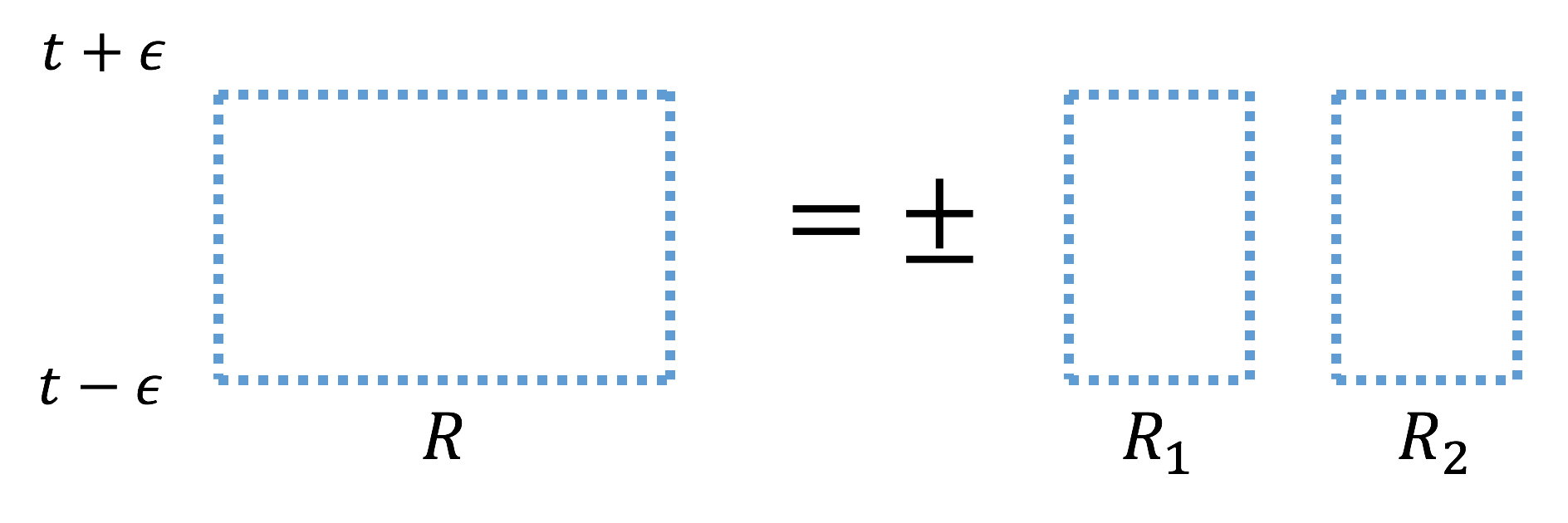}
\caption{The splitting of the symmetry  line $U_g(M)$ (shown in dotted blue lines) of a $\bZ_2$ global symmetry in 2d, where $M$ is the boundary of the rectangle $R\times [t-\epsilon,t+\epsilon]$.  This splitting obeys \eqref{splittable}. For a non-anomalous (anomalous) $\bZ_2$ global symmetry, we have a $+$ ($-$) sign on the right-hand side \cite{Chang:2018iay,Lin:2019kpn,Cordova:2019wpi}. The $\bZ_2$ symmetry of the Ising model is non-anomalous and therefore the sign above is $+$. }\label{fig:split}
\end{figure}

To extend the argument of \cite{Harlow:2018tng} to the more general topological operators, we ask: \textit{are the general topological operators/defects splittable?}\footnote{ In  \cite{Harlow:2018tng}, the splittability condition is stated for symmetry operators $U_g(R)$ at a fixed time on a disk $R$ in space.  However, symmetry operators/defects generally \textit{cannot} be defined on a manifold with boundary  in spacetime (see Section \ref{sec:endable}).  For example, in the $\bZ_N$ gauge theory in two spacetime dimensions, there is a $\bZ_N$ zero-form global symmetry whose symmetry operator is the  Wilson line $\exp[i  \oint A^{(1)}]$ (see Section \ref{ssec:ZN}).  
The Wilson line  can only be defined on a closed curve, not  on a finite segment   in spacetime in pure $\bZ_N$ gauge theory.  
For  a non-invertible topological operator/defect $T$,  even when $T(R)$ can be defined on such a disk  $R$, it does not have an inverse. 
 Nonetheless, one can still discuss the splittability of a topological  defect $T(M)$ on a closed manifold $M$ in the sense of \eqref{splittable}.}  
The short answer is no, at least not in the form  stated in \eqref{splittable}.

This can be seen from the non-invertible topological line $T$ in the 2d Ising CFT  associated with the Kramers-Wannier duality (see Section \ref{sec:noninvertible}).  
It obeys the splitting relation shown in Figure \ref{fig:nonsplit} (see, for example, Section 5.1.1 of   \cite{Chang:2018iay}), which does not take the form of \eqref{splittable}.  
Instead, the topological line $T(M)$ is equated to a superposition of two different terms, with nontrivial weighing coefficients $1/\sqrt{2}$. 
The last term on the right-hand side of Figure \ref{fig:nonsplit}  does not take the form of $T(R_1)T(R_2)$, but has an additional symmetry line joining between the two $T(R_i)$.   
This is to be contrasted with the corresponding splitting relation of a $\bZ_2$ symmetry line shown in Figure \ref{fig:split}, in which case \eqref{splittable} is obeyed.  
These novelties for the non-invertible topological operators/defects make an extension of the argument of \cite{Harlow:2018tng} nontrivial, but it is conceivable that a more general notion of splittability can be defined for topological operators, enabling a proof of the conjecture. 
We leave this for future studies.

\section{Conclusions}\label{sec:CONC}

We have seen that the standard lore that completeness of the spectrum of $G$ gauge theory follows from the absence of one-form global symmetries must be modified when $G$ is a finite (non-abelian)   group. Rather, completeness of the spectrum is equivalent to the absence of topological Gukov-Witten operators. We have proven this statement in three spacetime dimensions, and (under certain assumptions) we have shown that this proof generalizes to four or more dimensions.

It is natural to conjecture that quantum gravity does not have topological operators, extending the usual statement that quantum gravity does not permit exact global symmetries. Several lines of evidence suggest this conjecture is true, but neither the black hole argument against continuous global symmetries \cite{Misner, Banks:2010zn} nor the holographic argument against global symmetries in AdS$_d$ \cite{Harlow:2018tng} immediately extend to more general topological operators. It would therefore be worthwhile to formulate a more general proof against topological operators in quantum gravity.

In this work, we have dealt with the question of \emph{exact} global symmetries and \emph{exactly} topological operators, whose correlations functions are invariant under small deformations on the manifolds on which they are supported. From a phenomenological perspective, the distinction between an exactly topological operator and an approximately topological operator (which has only slight sensitivity to deformations of the manifold on which it is supported) is irrelevant provided the effects of such deformations are small enough to avoid experimental detection. 
Several works have sought to address this problem by investigating the strength of global symmetry-violating effects in quantum gravity  \cite{Fichet:2019ugl, Nomura:2019qps, Daus:2020vtf,approximatesym}; it would be nice to extend these analyses to more general topological operators and to study the phenomenological implications thereof.

The absence of global symmetries and the completeness hypothesis are two examples of Swampland conjectures that can be understood as consequences of the absence of topological operators in quantum gravity. It would be interesting to see what other Swampland statements may be connected to this web of ideas.

\section*{Acknowledgements}

We thank Maissam Barkeshli, Michele Del Zotto, Daniel Harlow, Ben Heidenreich, Po-Shen Hsin, Luca Iliesiu, Wenjie Ji, Theo Johnson-Freyd, Ho Tat Lam, Jacob McNamara, Miguel Montero, Hirosi Ooguri, Matthew Reece, Nathan Seiberg, Irene Valenzuela, and Qing-Rui Wang  for helpful discussions. We thank Daniel Harlow, Ben Heidenreich, Jacob McNamara, Miguel Montero, Hirosi Ooguri, Matthew Reece, and Irene Valenzuela for comments on the manuscript. The work of TR\ is supported by the Roger Dashen Membership and by NSF grant PHY-1911298. 
The work of SHS is supported by the Simons Foundation/SFARI (651444, NS).

\appendix

\newpage

\bibliographystyle{utphys}
\bibliography{ref}

\end{document}